\pdfoutput=1

\documentclass[10pt,journal,compsoc]{IEEEtran}
\ifCLASSOPTIONcompsoc
  \usepackage[nocompress]{cite}
\else
  \usepackage{cite}
\fi
\ifCLASSINFOpdf
  \usepackage[pdftex]{graphicx}
  \graphicspath{{program/output/figures/}}
  \DeclareGraphicsExtensions{.pdf,.png}
\else
  \usepackage[dvips]{graphicx}
  \graphicspath{{program/output/figures/}}
  \DeclareGraphicsExtensions{.eps}
\fi
\usepackage{url}

\usepackage{hyperref}
\usepackage[capitalize]{cleveref}
\usepackage{color}
\usepackage{gensymb}
\usepackage{ifthen}
\usepackage[percent]{overpic}
\usepackage{stackengine}

\newcommand{\cf}{\emph{cf}~}
\newcommand{\eg}{\emph{e.g.}~}
\newcommand{\ie}{\emph{i.e.}~}
\newcommand{\vs}{\emph{vs}~}
\newcommand{\AuthorName}{Eric Bruneton}

\newcommand{\supplemental}[1]{}

\newlength{\mytablewidth}
\newenvironment{mytabular}{%
  \setlength{\tabcolsep}{0cm}
  \renewcommand\arraystretch{0}
  \tabular%
}{%
  \endtabular
}

\newcommand{\myfigure}[2]{%
\begin{figure*}\begin{center}#1\end{center}#2\end{figure*}%
}

\newlength\lunderset
\newlength\rulethick
\def\stackalignment{l}
\newcommand\nunderline[2][1]{\setbox0=\hbox{#2}%
  \stackunder[#1\lunderset-\rulethick]{\strut#2}{\rule{\wd0}{\rulethick}}}

\begin{document}
%
\title{A Qualitative and Quantitative Evaluation of\\ 8 Clear Sky Models}
%
%
%
%

\author{\AuthorName\thanks{\copyright 2016 IEEE. Personal use is permitted. For
any other purposes, permission must be obtained from the IEEE by emailing
pubs-permissions@ieee.org. This is the author's version of an article that has
been published in this journal. Changes were made to this version by the
publisher prior to publication. The final version of record is available at 
http://dx.doi.org/10.1109/TVCG.2016.2622272.}}

%
%

\markboth{IEEE Transactions on Visualization and Computer Graphics}%
{Bruneton: A Qualitative and Quantitative Evaluation of 8 Clear Sky Models}
%


\IEEEtitleabstractindextext{%
\begin{abstract}
We provide a qualitative and quantitative evaluation of 8 clear sky models used
in Computer Graphics. We compare the models with each other as well as with
measurements and with a reference model from the physics community. After a
short summary of the physics of the problem, we present the measurements and the
reference model, and how we "invert" it to get the model parameters. We then
give an overview of each CG model, and detail its scope, its algorithmic
complexity, and its results using the same parameters as in the reference model.
We also compare the models with a perceptual study. Our quantitative results
confirm that the less simplifications and approximations are used to solve the
physical equations, the more accurate are the results. We conclude with a
discussion of the advantages and drawbacks of each model, and how to further
improve their accuracy.
\end{abstract}

}

\maketitle

\IEEEdisplaynontitleabstractindextext

%
\IEEEpeerreviewmaketitle

\IEEEraisesectionheading{\section{Introduction}\label{sec_introduction}}

%
%
%
%
\IEEEPARstart{T}{he} sky is a key element in outdoor scenes, and also the
primary light source for many indoor scenes, and thus an important topic in
Computer Graphics. The physical equations that explain the sky color are well
understood, but they are computationally expensive to solve. This motivated the
design of efficient and accurate clear sky rendering algorithms in the Computer
Graphics community.

In this paper we evaluate qualitatively and quantitatively 8 of these models,
namely the Nishita93\cite{Nishita93}, Nishita96\cite{Nishita96},
Preetham\cite{Preetham99}, O'Neal\cite{ONeal05}, Haber\cite{Haber05},
Bruneton\cite{Bruneton08}, Elek\cite{Elek10} and Hosek\cite{Hosek12} models. Our
quantitative evaluation is based on ground-truth measurements of a real clear
sky made by Kider et al.\cite{Kider14}. We also compare the Computer Graphics
models with libRadtran\cite{Mayer05}, a well known, thoroughly tested model used
in hundreds of publications in atmospheric sciences. For this:
\begin{itemize}
\item we estimate the model parameters, such as the density of the aerosols and
their properties, by finding the values yielding the best fit of the
ground-truth measurements by the libRadtran results,
\item we use these parameters as input to the models to evaluate, and compare
their results with the measurements and with the libRadtran results. We also
compare these results with a perceptual study.
\end{itemize}
The rest of this paper is organized as follows: after some related work in
Section~\ref{sec_relatedwork}, we present the reference measurements and the
reference model in Section~\ref{sec_reference}. We continue with our comparison
framework in Section~\ref{sec_framework} and our comparison results for each
model in Sections~\ref{sec_nishita93} to \ref{sec_hosek}.
Section~\ref{sec_study} presents our perceptual study. We conclude with a
discussion of the advantages and drawbacks of each model, and of possible
methods to improve their accuracy, in Section~\ref{sec_discussion}.

\section{Related work}\label{sec_relatedwork}

Very few papers have been published on the evaluation and comparison of the
clear sky models in Computer Graphics. Sloup\cite{Sloup02} provides a survey
with some qualitative comparisons, but this work does not include some of the
more recent models and does not contain quantitative comparisons. A more recent
survey can be found in\cite{Lopes14}. It provides a few more qualitative
comparisons (including a comparison of the rendering results) and a comparison
of the frame rate obtained with each model. The closest paper to our work is the
paper from Kider et al.\cite{Kider14}, which provides ground-truth measurements
and a quantitative comparison of six clear sky models.

Our work is based on the ground truth measurements from Kider et al., and
extends their clear sky model comparisons with:
\begin{itemize}
\item two more Computer Graphics models: O'Neal\cite{ONeal05} and
Elek\cite{Elek10},
\item a reference model from the physics community: libRadtran\cite{Mayer05}
(version 2.0.1),
\item a comparison of the absolute and relative luminance of the models, and of
their chromaticity,
\item a comparison of the time and memory complexity of each model,
\item a comparison of the scope and limitations of each model (\eg supported
viewpoints and sun angles, support of aerial perspective or not, etc),
\item a perceptual study of the rendering results of each model,
\item the full source
code\footnote{\url{https://github.com/ebruneton/clear-sky-models}} of our
implementation of these models.
\end{itemize}

\section{Reference measurements and model}\label{sec_reference}

This section presents the reference measurements and model used to
quantitatively evaluate the 8 Computer Graphics models in the next sections. We
first present the measurements, then the model, and finally the model
parameters.

\subsection{Measurements}

In order to compare the models against the ground truth, we use the measurements
provided in Kider et al.\cite{Kider14}. These measurements consist of 81 samples
over the sky dome, for 17 daytimes between 9h30 and 13h30. Each sample is a
radiance spectrum covering all the wavelengths from 350 to 2500 $nm$, in steps
of 1 $nm$. The dataset also includes irradiance measurements and HDR photos.
See\cite{Kider14} for more details about the acquisition of this data.

\subsection{Physical model}

The sky color results from the scattering and absorption of the sun light by the
air molecules and the aerosol particles. A full presentation and discussion of
the underlying physical equations can be found in\cite{Sloup02} and in the
papers corresponding to the 8 models evaluated here. In this section we only
provide a brief overview.

The scattering of light by air molecules is described by the Rayleigh theory. It
is proportional to $\lambda^{-4}$, where $\lambda$ is the wavelength, which
explains the blue color of the sky. The scattering phase function, which
describes the directional dependence of the scattering, is smooth and
independent of the wavelength. The scattering is also proportional to the
density of the molecules, which does not vary much with the weather conditions
and is approximately exponentially decreasing with the altitude. Finally, the
air molecules also absorb some light (\eg ozone), but mostly outside the visible
spectrum (in the 360-830 $nm$ range air molecules absorb only 1.5\% of the
light).

In contrast, aerosols scatter and absorb light in a much more complex and highly
variable way. The scattering is described by the Mie theory, and depends on the
probability distribution function of the particle sizes. The resulting phase
function is strongly anisotropic, usually with a strong forward peak, and can
depend on wavelength and on altitude. The aerosols also absorb a non-negligible
amount of light in the visible range, and this absorption depends on the
wavelength. Finally the density of the aerosol particles and its height
dependence are highly variable, from clean to polluted conditions.

The ground also plays a role in the sky color\cite{Hosek12}. Its albedo and BRDF
can vary a lot, \eg between ocean, forests, sand or snow conditions.

\subsection{Reference solver}

libRadtran\cite{Mayer05} is widely used in atmospheric science, and cited in
hundreds of
papers\footnote{\url{http://www.libradtran.org/doku.php?id=publications}}. It
provides a unified way to specify the atmospheric conditions, which can then be
solved to get the sky radiance (amongst other quantities), using a variety of
solvers. The default solver is DISORT\cite{Stamnes88}, based on the discrete
ordinates algorithm.

The input parameters are specified by layers. Any number of layers can be used,
and for each layer the pressure, temperature, density of air molecules and
absorption by air molecules can be specified. For aerosols, each layer can
specify wavelength-dependent scattering and absorption coefficients, as well as
a wavelength-dependent phase function. It is also possible to specify water or
ice clouds. Finally, the extraterrestrial sun radiance spectrum and the ground
spectral albedo and BRDF can be freely specified. Sophisticated BRDF models for
vegetation and for the ocean (depending for instance on wind speed) are
provided.

Once a solver is chosen (different options are available, such a plane-parallel,
pseudo-spherical or fully spherical geometry, support of polarization or not,
etc), the sky radiance can be obtained for any number of view directions and any
number of wavelengths or wavelength bands.

\subsection{Model parameters}\label{sec_model_parameters}

In order to compare the Computer Graphics models with the ground truth, we need
values for all the above parameters (e.g. the scattering and extinction
coefficients, density profiles, Mie phase function and ground BRDF).

Ideally we would use direct measurements for these parameters, but we don't have
such data. Column integrated values (as opposed to per layer values) could be
inferred by model inversion from dense radiance measurements in the solar
aureole region\cite{Dubovik00}. But we don't have this data either. Some
inversion results are provided by the AERONET project\cite{Holben98}, for about
1000 ground stations over the world, since 1993. However, the nearest data
corresponding to the Kider measurements, made at the Cornell University, is from
a ground station near Toronto.

Thus, due to the lack of direct or indirect measurements of the model
parameters, we decided to compute them by "inverting" the libRadtran model. For
this we computed the parameter values that minimize the root mean square error
(RMSE) between the libRadtran results and the measurements (summed over all the
radiance measurement samples). Note that this method does not bias our results,
since we don't use the CG models we want to evaluate to compute their parameters
(and since we don't evaluate libRadtran here).

\subsubsection*{Inversion method}

To reduce the complexity of this nonlinear optimization problem, we first
eliminated as many parameters as possible. For this:
\begin{itemize}
\item we used the parameter space of the CG models, smaller than the full
parameter space. Indeed, there is no point in computing parameters that are
ignored by these models. We thus assumed that air molecules do not absorb light,
that the aerosols phase function does not depend on wavelength, that the ground
has a Lambertian BRDF, etc.
\item we chose typical values for the air molecules parameters, which as
discussed above do not vary much: scattering coefficient at sea level
from\cite{Penndorf57} at $15\,^\circ{\mathrm C}$, with an exponentially
decreasing value using a scale height of $8\,km$.
\item we fixed the ground albedo to the grass spectral albedo from\cite{Uwe95}
(measurements were made in the New-York state in May which, at this time of the
year, is mostly covered by vegetation).
\item we fixed the scale height for the aerosol particles to $1.2\,km$, a value
commonly used, and their single scattering albedo (ratio of scattering to
extinction coefficient) to $0.8$, independent of wavelength.
\end{itemize}
Thanks to these simplifications, the only remaining free parameters are the
optical depth and the phase function of the aerosols. Assuming an aerosol
optical depth of the form $\beta\lambda^{-\alpha}$ (\ie using the \r{A}ngstr\"om
turbidity formula\cite{Angstrom61}, with $\lambda$ in $\mu m$) and a
Cornette-Shanks\cite{TS99} phase function depending on a single asymmetry
parameter $g$, this translates to 3 free dimensionless parameters: $\alpha$,
$\beta$ and $g$.

\subsubsection*{Inversion result}

Using $11\times10\times5$ uniformly distributed samples in the
$[0,2]_{\alpha}\times[0.02,0.2]_{\beta}\times[0.5,0.9]_g$ parameter space (which
covers a wide range of aerosol properties, from very clear to hazy conditions),
we found a minimum RMSE at $0.8,0.04,0.7$: $11.48\ mW/(m^2.sr.nm)$. This result
can be refined with NLopt\cite{NLopt} to $0.816,0.0384,0.704$
($\mathrm{RMSE}=11.45$). However, due to the approximations made above when
eliminating some parameters, and because the RMSE is nearly constant in a large
ellipsoid in the $[0.6,0.9]\times[0.038,0.042]\times[0.68,0.73]$ region (see the
supplemental material), this refinement does not really give any additional
significant digits. Thus, in the following, we used $\alpha=0.8$, $\beta=0.04$
and $g=0.7$.

\subsubsection*{Turbidity}

In addition to the above parameters, we also need the value of the turbidity
$T$, for the Preetham and Hosek models. For this we used the turbidity that
minimizes the RMSE between the measurements of the zenith luminance $L_z$ and
the empirical relation
$L_z=(1.376T-1.81)\cot\theta_{sun}+0.38\,kcd/m^2$\cite{Karayel84}, namely
$T=2.53$. This only gives a coarse estimate of the turbidity, but our
qualitative results and models ranking stay unchanged for any value of $T$
between $2$ and $3$ (see the supplemental material), and in particular for the
values that minimizes the RMSE between the Preetham (resp. Hosek) model and the
measurements ($T=2.33$, RMSE$_{\mathrm{min}}=86$ - resp. $T=2.62$,
RMSE$_{\mathrm{min}}=41.45$).

\section{Comparison framework}\label{sec_framework}

Before evaluating the Computer Graphics models with the measurements and with
the reference model, in the next sections, we present here how we evaluate them
qualitatively and quantitatively.

\subsection{Qualitative evaluations}

Our qualitative evaluations include a scope and algorithmic complexity analysis.
The scope is defined by the viewpoints and sun directions supported by the
model. Some models support only views from the ground level, while others
support any viewpoint from ground to space. Some models can only compute the sky
color, while others support aerial perspective, required to render realistic
terrains. Finally, some models are limited to sun directions above the horizon,
while others support any sun direction.

The algorithmic complexity is given by the time and memory complexity of the
precomputation and rendering phases. All the models have a polynomial complexity
and, to facilitate comparisons, we give it in terms of a single abstract
parameter $n$. A complexity $O(n^k)$ means that if one uses 2 times more samples
in each single numeric integration\footnote{We count integrals over directions
as double integrals.} and along each array dimension, the complexity increases
by a factor $2^k$. All the models also have a linear complexity in the number of
wavelengths $n_\lambda$, which is omitted in the following -- \ie $O(n^k)$ is a
shortcut for $O(n_\lambda n^k)$.

\subsection{Quantitative evaluations}\label{sec_quantitativeeval}

In order to compare the models with each other in a sound way we use the same
atmospheric parameters for each model, when applicable. These parameters include
the extraterrestrial solar spectrum, the ground albedo and the Rayleigh and Mie
scattering and absorption coefficients, phase functions, density profiles, etc.

\begin{figure*}
\newcommand{\hour}{11h45}
\begin{center}
\includegraphics[width=\columnwidth]{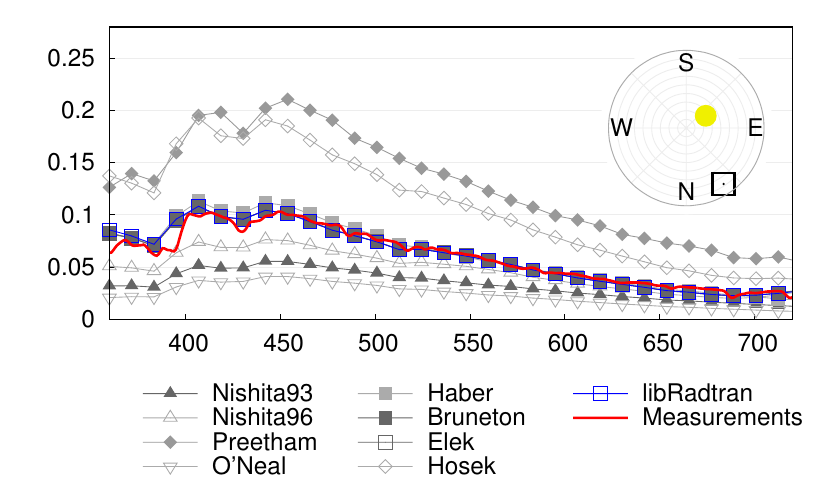}
\includegraphics[width=\columnwidth]{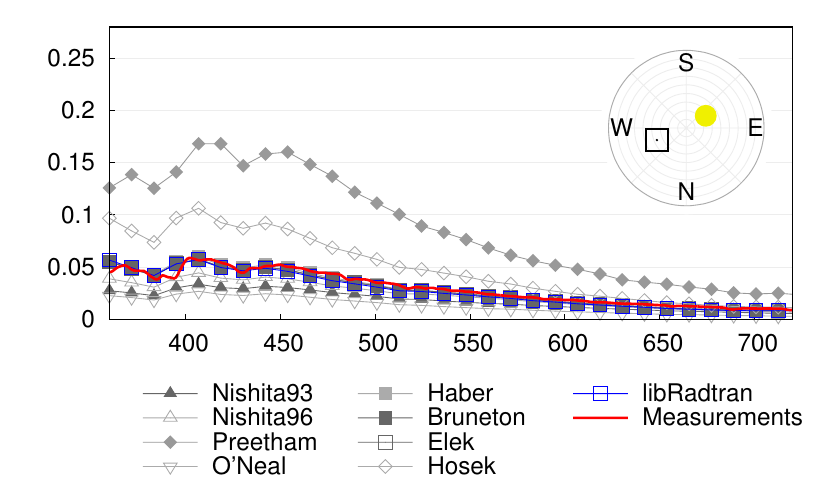}\vspace{-1.3cm}
\includegraphics[width=\columnwidth]{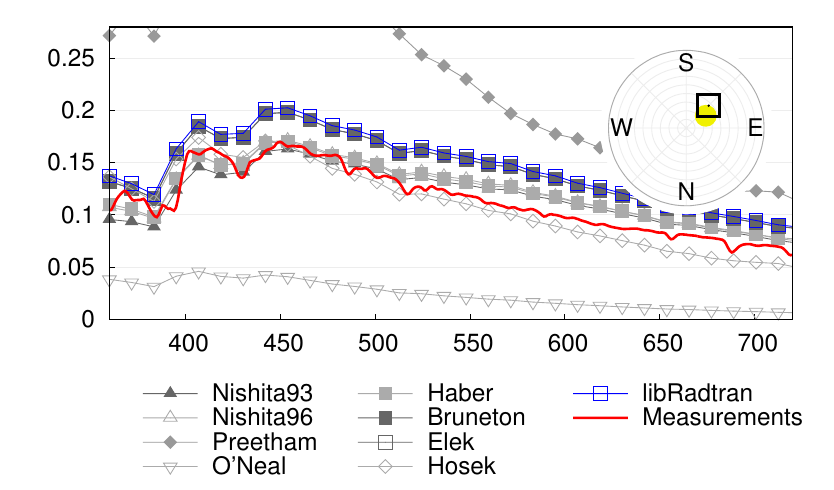}
\includegraphics[width=\columnwidth]{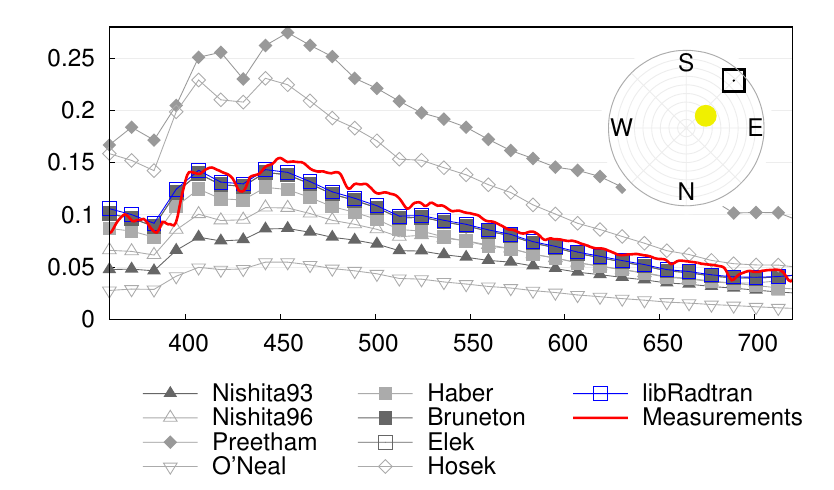}\vspace{-0.5cm}
\end{center}
\caption{{\bf Spectral radiance}. Four spectral radiance samples in
$W.m^{-2}.sr^{-1}.nm^{-1}$ as a function of wavelength in $nm$, for
$t=\mathrm{\hour}$ (sun zenith angle =
\protect\input{program/output/figures/sza_\hour.txt}\unskip\degree). Compare
with Fig. 10 in\cite{Kider14}.}\label{fig:spectral_radiance}
\end{figure*}

\begin{figure}
\begin{center}
\includegraphics[width=\columnwidth]{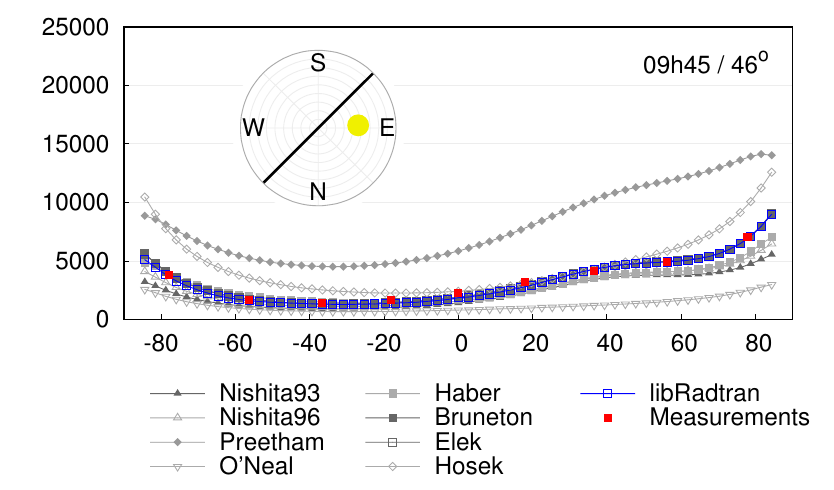}\vspace{-1.3cm}
\includegraphics[width=\columnwidth]{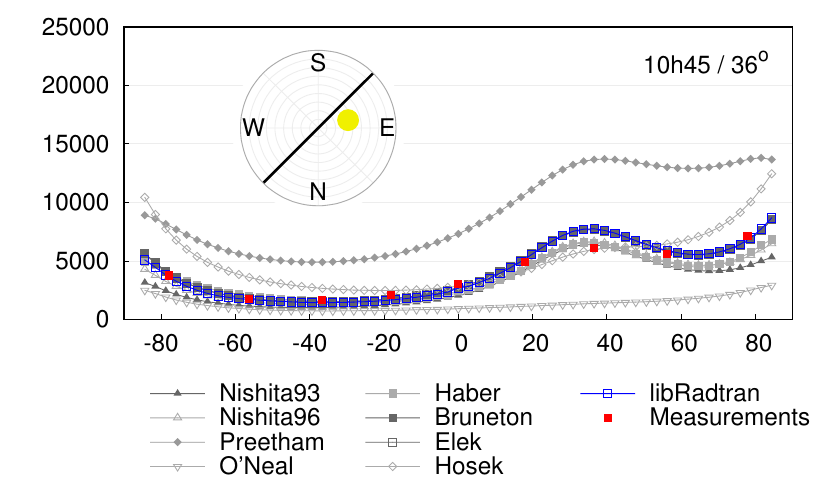}\vspace{-1.3cm}
\includegraphics[width=\columnwidth]{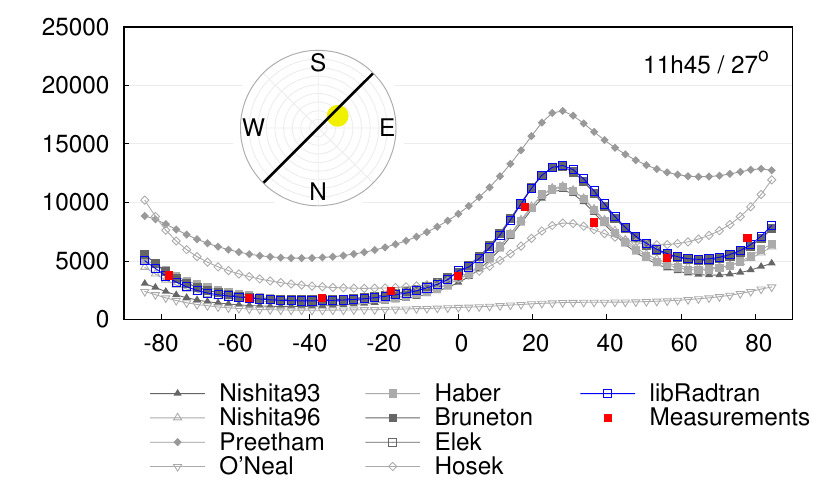}\vspace{-1.3cm}
\includegraphics[width=\columnwidth]{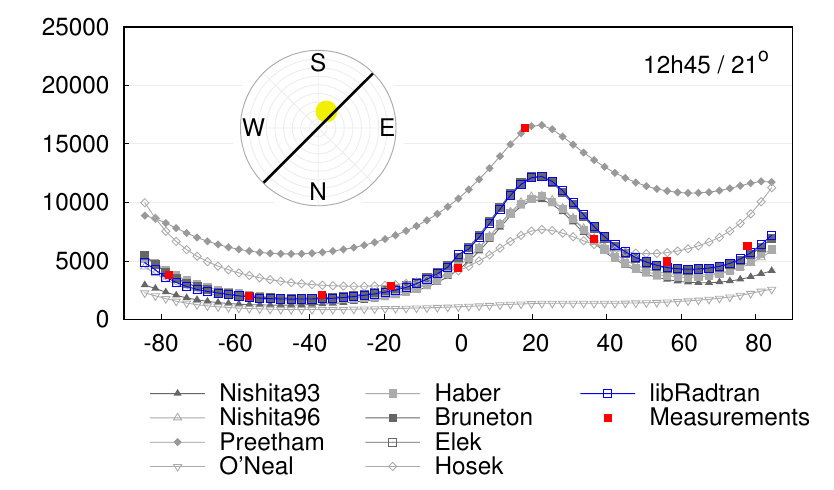}\vspace{-0.5cm}
\end{center}
\caption{{\bf Luminance profiles}. The sky luminance in $cd.m^{-2}$ in a fixed
vertical plane (black line), as a function the view zenith angle, for different
time of day / sun zenith angle values.}\label{fig:luminance_profile}
\end{figure}

Likewise, and to facilitate comparisons with the results from Kider et al., we
run each model spectrally, using the same 40 wavelengths between $360\,nm$ and
$830\,nm$ as in\cite{Kider14} (although the models originally use between 3 and
15 wavelengths -- we discuss the impact of the number of wavelengths used in
\cref{sec_rgborspectral}). We then plot the results obtained with each model in
different forms:
\begin{itemize}
\item a rendering of the skydome, in \cref{fig:rendering},
\item the absolute luminance, in \cref{fig:absolute_luminance},
\item the relative luminance, in \cref{fig:relative_luminance},
\item the chromaticity, in \cref{fig:chromaticity},
\item the relative error with the measurements, in \cref{fig:relative_error},
\item the spectral radiance for 4 samples, in \cref{fig:spectral_radiance},
\item the luminance in a fixed vertical plane, in \cref{fig:luminance_profile},
\item the sky irradiance as a function of time, in \cref{fig:sky_irradiance}.
\end{itemize}

The next sections present the 8 Computer Graphics models we want to evaluate, in
chronological order. Each section gives a short overview of the model, some
important implementation details if applicable, and then our qualitative and
quantitative results.

\section{Nishita93 Model}\label{sec_nishita93}

\subsection{Overview}

The Nishita93 model\cite{Nishita93} is one of the first realistic sky rendering
algorithms in Computer Graphics. It is based on a numerical integration of the
single scattering equation (and thus neglects multiple scattering). For each
pixel, an integral along the corresponding view ray is computed numerically. The
integrand includes two transmittance terms: one from the viewer to the current
sample along the view ray, and one from this sample to the sun. Both terms
require a numerical integration. To avoid having to compute a double integral at
each pixel, Nishita et al. compute the first term incrementally, while
evaluating the outer integral, and precompute the second term in a 2D array.

\newcommand{\modelrow}[2]{
\rotatebox{90}{\footnotesize #2 /
\input{program/output/figures/sza_#2.txt}\unskip\degree} &
\includegraphics[width=0.198\mytablewidth]{#1_#2_nishita93} &
\includegraphics[width=0.198\mytablewidth]{#1_#2_nishita96} &
\includegraphics[width=0.198\mytablewidth]{#1_#2_preetham} &
\includegraphics[width=0.198\mytablewidth]{#1_#2_oneal} &
\includegraphics[width=0.198\mytablewidth]{#1_#2_haber} &
\includegraphics[width=0.198\mytablewidth]{#1_#2_bruneton} &
\includegraphics[width=0.198\mytablewidth]{#1_#2_elek} &
\includegraphics[width=0.198\mytablewidth]{#1_#2_hosek} &
\includegraphics[width=0.198\mytablewidth]{#1_#2_libradtran} &
\ifthenelse{\equal{#2}{06h00} \OR \equal{#2}{07h00} \OR \equal{#2}{08h00}}
{\raisebox{0.091\mytablewidth}{\makebox[0.198\mytablewidth]{\footnotesize n/a}}}
{\includegraphics[width=0.198\mytablewidth]{#1_#2_measurements}}}

\newcommand{\modeltable}[1]{
\begin{mytabular}{p{0.04\mytablewidth}*{10}{p{0.198\mytablewidth}}}
&
\footnotesize\bfseries Nishita93 &
\footnotesize\bfseries Nishita96 &
\footnotesize\bfseries Preetham &
\footnotesize\bfseries O'Neal &
\footnotesize\bfseries Haber &
\footnotesize\bfseries Bruneton &
\footnotesize\bfseries Elek &
\footnotesize\bfseries Hosek &
\footnotesize\bfseries\leavevmode\color{blue} libRadtran &
\footnotesize\bfseries\leavevmode\color{red} Measurements\vspace{1mm}\\
\modelrow{#1}{06h00}\\
\supplemental{\modelrow{#1}{07h00}\\}
\supplemental{\modelrow{#1}{08h00}\\}
\supplemental{\modelrow{#1}{09h30}\\}
\supplemental{\modelrow{#1}{09h45}\\}
\supplemental{\modelrow{#1}{10h00}\\}
\modelrow{#1}{10h15}\\
\supplemental{\modelrow{#1}{10h30}\\}
\supplemental{\modelrow{#1}{10h45}\\}
\supplemental{\modelrow{#1}{11h00}\\}
\modelrow{#1}{11h15}\\
\supplemental{\modelrow{#1}{11h30}\\}
\supplemental{\modelrow{#1}{11h45}\\}
\supplemental{\modelrow{#1}{12h00}\\}
\supplemental{\modelrow{#1}{12h15}\\}
\supplemental{\modelrow{#1}{12h30}\\}
\supplemental{\modelrow{#1}{12h45}\\}
\supplemental{\modelrow{#1}{13h00}\\}
\modelrow{#1}{13h15}\supplemental{\\}%
\supplemental{\modelrow{#1}{13h30}}
\end{mytabular}}

\myfigure{\modeltable{image_full_spectral}}{%
\caption{{\bf Rendering}. Fisheye skydome rendering of the spectral radiance
obtained with each model, convolved with the CIE color matching functions,
converted from XYZ to linear sRGB, and tone mapped with $1-e^{-kL}$, for several
time of day / sun zenith angle values (the red cross indicates the sun
direction). The measurements are interpolated using bicubic spherical
interpolation before rendering. Compare with Fig. 13
in\cite{Kider14}.\label{fig:rendering}}}

\subsection{Our Implementation}

The Nishita93 model discretizes the atmosphere in concentric layers of constant
optical depths (for better precision compared to layers of constant heights),
and uses the intersections of these spherical layers with concentric cylinders
oriented towards the sun as sampling points for the precomputed 2D array of
transmittances. The number of layers and cylinders is not specified
in\cite{Nishita93}. In our implementation, we use $n_s=64$ spheres (\ie 63
layers) and $n_c=64$ cylinders (doubling $n_s$ and $n_c$ decreases the RMSE by
less than 2\%).

\subsection{Qualitative evaluation}

Although this model uses some precomputed data, this data does not depend on the
viewer position nor on the sun direction. The model therefore supports all
viewpoints from ground to space, aerial perspective (simply by integrating
single scattering between the viewer and the terrain), and sun directions below
the horizon.

The precomputation phase evaluates a single integral with $O(n_s)$ samples for
all elements of a $n_s\times n_c$ array, and therefore has $O(n_s^2n_c)$ and
$O(n_s n_c)$ time and memory complexity, respectively. Thanks to this
precomputed table, rendering a pixel has $O(n_s)$ time complexity (instead of
$O(n_s^2)$ without).

\subsection{Quantitative evaluation}

\newcommand{\quantitativefigs}{\cref{%
fig:rendering,%
fig:absolute_luminance,%
fig:relative_luminance,%
fig:chromaticity,%
fig:relative_error,%
fig:luminance_profile,%
fig:spectral_radiance,%
fig:sky_irradiance}}

Our quantitative results, in \quantitativefigs, show that this model almost
always underestimates the measured values. For instance, computed irradiances
(\cref{fig:sky_irradiance}) are about one third less than the measured values.
This seems logical since multiple scattering is ignored in this model. This also
indicates that, in the measured values, multiple scattering is responsible for
one third of the sky irradiance, which seems consistent with\cite{Bary74}
(especially the values from Table 1, for a turbidity $T=2$). The RMSE between
the Nishita93 model and the measurements, summed over the 17 daytimes, 81
direction samples, and the wavelengths between $360$ and $720\,nm$, is
$26.6
\ mW/(m^2.sr.nm)$.

The relative luminance (\cref{fig:relative_luminance}) is quite similar to the
measurements, but the chromaticity (\cref{fig:chromaticity}) shows some
differences with the libRadtran results near the horizon and for sunrise and
sunset (not measured).

\section{Nishita96 Model}\label{sec_nishita96}

\subsection{Overview}\label{sec_nishita96overview}

The Nishita96 model\cite{Nishita96} adds (approximate) multiple scattering to
the Nishita93 model. The model supports any scattering order, but the authors
compute only double scattering. Computing double scattering involves the
computation of an integral over all directions, at each sample point along the
view ray, with an integrand equal to the single scattering integral. In other
words, it requires an additional triple integral at each sample along the view
ray. To optimize this computation, Nishita et al. propose to:
\begin{itemize}
\item reduce the integral over all directions to a sum over 8 specific
directions, called {\em sampling} directions, in the plane containing the
vertical and sun vectors.
\item for each of these 8 directions, precompute the single scattering at the
vertices of a $n_x\times n_y \times n_z$ 3D grid having one axis parallel to
this direction. Single scattering is computed incrementally, from one vertex to
the next along this axis.
\end{itemize}
Then, at rendering time, 8 trilinearly interpolated lookups into these 8
precomputed tables, at each sample point along the view ray, are sufficient to
compute the contribution from double scattering.

\myfigure{\modeltable{absolute_luminance}

\includegraphics[width=1.5\mytablewidth]{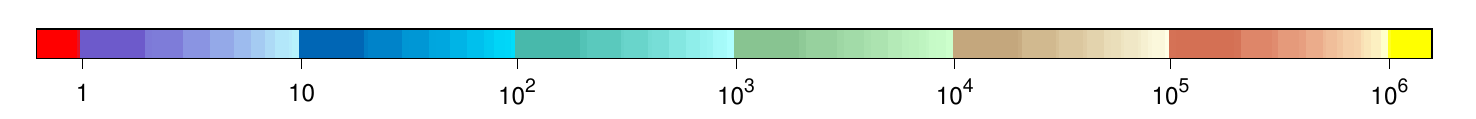}\vspace{-0.5cm}}{%
\caption{{\bf Absolute luminance}. The absolute luminance in $cd.m^{-2}$
obtained with each model, using the same color scale as in Fig.~7
in~\cite{Zotti07}. The measurements are interpolated using bicubic spherical
interpolation before being converted to luminance
values.\label{fig:absolute_luminance}}}

\subsection{Our Implementation}

In our implementation we use $32^3$ voxels for each of the 8 precomputed tables
for single scattering. The voxels cover the whole sky dome (\ie about
$1700 \times 1700 \times 60$ $km$) using 2 orthogonal horizontal axes and a
third axis aligned with the sampling direction (or the vertical if the sampling
direction is horizontal). We also ignored the third and higher orders of
scattering, as in\cite{Nishita96}, but we did not use Eq. (5) of\cite{Nishita96}
to compute double scattering. Indeed, this equation seems incorrect to us (one
parenthesis is missing, and we couldn't find a way to get a homogeneous equation
by adding it back). Instead, we used the following equation (using the notations
from\cite{Nishita96} and omitting $\lambda$ parameters):
\begin{eqnarray*}
I_v=I_s\int_{P_v}^{P_a}e^{-t(s)}\Big[R(s,\theta)e^{-t(s')}
+\int_{4\pi}R(s,\theta_w)&\\
\int_P^{P_w}R(s_w,\theta'_w)e^{-t(s_w)-t(s'_w)}\mathrm{d}s_w\mathrm{d}w\Big]
\mathrm{d}s
\end{eqnarray*}
We validated it by comparing its single and double scattering terms with those
of the Haber and Bruneton models, which yield similar results (see the
supplemental material).

\subsection{Qualitative evaluation}

The 8 sampling directions are the zenith, the sun direction, the opposites, and
the orthogonal, in the same plane, to these four directions. Hence, 4
precomputed tables must be recomputed for each new sun direction\footnote{It
would also be possible to precompute $4n$ tables for $n$ sun directions and use
quadrilinear interpolation at rendering time, but this is not in the original
model.}. This prevents this model to be used for viewpoints in space, where
points with different local sun zenith angles can be viewed simultaneously. On
the other hand, aerial perspective and arbitrary sun directions are supported,
as in Nishita93.

The algorithmic complexity is dominated by the multiple scattering computations.
Assuming we keep 8 sampling directions in all cases, precomputing the single
scattering 3D tables has a $O(n_x n_y n_z)$ complexity both in time and memory
(the time complexity would be $O(n_x n_y n_z^2)$ without the incremental
computation). Rendering a pixel has the same $O(n_z)$ time complexity as
Nishita93, but with a larger constant factor ($n_z$ is equivalent to $n_s$, and
$n_x$ and $n_y$ to $n_c$).

\subsection{Quantitative evaluation}

Our quantitative results show that the addition of double scattering improves
the accuracy of the model, compared to Nishita93. In particular, the RMSE
decreases from
26.6
\unskip\ for Nishita93
to 18.3
\unskip\ for
Nishita96. However, this model still underestimates the measured values. This is
probably due to both the approximate double scattering evaluation (using only 8
directions in a single plane instead of many directions over the whole unit
sphere), and the fact that third and higher orders of scattering are not taken
into account (since, as we show in the next sections, the models that do not use
these approximations are more accurate).

The relative luminance is a bit closer to the measurements, compared to
Nishita93, but the chromaticity remains quite different from the libRadtran
results near the horizon and for sunrise and sunset (not measured).

\section{Preetham Model}\label{sec_preetham}

\subsection{Overview}

The Preetham model\cite{Preetham99} is an analytical model allowing the
computation of the sky color or spectral radiance with a simple mathematical
formula. It was obtained by computing sky radiances for many view directions,
sun directions, and turbidity values, using the Nishita96 model, and then by
fitting the results with an analytical function from Perez et al.\cite{Perez93},
using non-linear least square fitting. The resulting model has only one
parameter, namely the turbidity. This model is completed with an independent
equation for aerial perspective, derived using a flat Earth hypothesis, and with
a model for the Sun radiance.

\myfigure{\modeltable{relative_luminance}

\includegraphics[width=1.5\mytablewidth]{scale_caption}\vspace{-0.5cm}}{%
\caption{{\bf Relative luminance}. The luminance obtained with each model, in
percentage of the zenith luminance, and using the same color scale as in Fig.~6
in~\cite{Zotti07}. The measurements are interpolated using bicubic spherical
interpolation before being converted to luminance
values.\label{fig:relative_luminance}}}

\subsection{Our implementation}

Our implementation is a straightforward implementation of the equations
from\cite{Preetham99}. Since the Preetham model has no input parameters for the
extraterrestrial solar spectrum, the ground albedo, the phase functions, etc
(all already included in the model), we simply ignored these parameters when
running the Preetham model.

\subsection{Qualitative evaluation}

The Preetham model is limited to views from the ground, both because the
analytical functions were fitted for an observer at the ground level, and
because the separate aerial perspective model assumes a flat Earth hypothesis.
It is also limited to sun directions above the horizon. It supports aerial
perspective, but with a separate model from the sky model, which could give
visual inconsistencies between the two (near the horizon).

The time and memory complexity to render a pixel is simply $O(1)$. There is no
precomputation phase at all, if one simply wants to use the model as is.
However, if one wants to change some atmospheric parameter, or the ground
albedo, it is necessary to recompute the sky radiance for many view directions
and sun directions, and perform a new non linear fitting. Using the Nishita96
model, the time and memory complexity of this precomputation would be at least
$O(n_x n_y n_z n)$ (for $n$ sun directions).

\subsection{Quantitative evaluation}

Our quantitative results, which confirm those of Zotti et al. (compare
\cref{fig:absolute_luminance} with Fig. 7 in\cite{Zotti07} for $T=2.5$), show
that the Preetham model overestimates the measured values, by a large factor
(about 2; the RMSE is
88.1
\unskip). The relative
luminance and the chromaticity are also quite different from the measured ones
and from Nishita96 or libRadtran (especially near the Sun or the horizon).

This is surprising, since this model was fitted using results from the Nishita96
model, which slightly underestimates the measured values. In any case, we
believe that the family of analytical functions used by Preetham, fitted
directly to the measurements, or to results from libRadtran or another reference
model, could produce a much better fit of the measured values (\cf the Hosek
model in Section~\ref{sec_hosek}, which uses very similar analytical functions
and gets much better results).

\section{O'Neal Model}\label{sec_oneal}

\myfigure{\modeltable{chromaticity}}{%
\caption{{\bf Chromaticity}. The rg-chromaticity, $(r,g,b)/\mathrm{max}(r,g,b)$,
obtained with each model. The measurements are interpolated using bicubic
spherical interpolation before being converted to chromaticity
values.}\label{fig:chromaticity}}

\subsection{Overview}

The O'Neal model\cite{ONeal05} is one of the first atmospheric scattering models
running at interactive frame rates on GPU. As in Nishita93, on which it is
based, this model uses a numerical integration of the single scattering
equation, and neglects multiple scattering. This integration is performed per
vertex instead of per pixel, with the phase functions evaluated per pixel to
avoid loosing angular precision. The main difference to the Nishita93 model is
how transmittance is computed in the numerical integration loop: instead of
using a precomputed 2D table (which would require texture fetches from the
vertex shader), O'Neal uses an analytic approximation of this 2D function.

\subsection{Our Implementation}

Our implementation is based on the author's source code, with some
simplifications removed:
\begin{itemize}
\item we use two scale height values, one for Rayleigh and one for Mie, as in
all the other models (O'Neal uses the same value for both, which divides the
number of transmittance terms to evaluate by 2),
\item we use the Rayleigh phase function for air molecules (O'Neal lets the user
choose between this or an isotropic phase function -- \cf section 16.5.2
in\cite{ONeal05}).
\end{itemize}
We also replaced the analytic approximation of the transmittance proposed by
O'Neal, which is specific to the atmospheric parameters he used, with a generic
one based on\cite{Schuler12}. Finally, we used 4 samples, instead of 2 in the
original source code, to compute each single scattering integral. This is more
or less equivalent to using $n_s=4$ concentric spheres in Nishita93.

\subsection{Qualitative evaluation}

The O'Neal model has the same scope as the Nishita93 model, on which it is
based. In other words it supports all viewpoints from ground to space, aerial
perspective, as well as sun directions below the horizon. Since the precomputed
2D transmittance table is replaced with an analytic approximation there is no
precomputation phase, unlike in Nishita93. But this does not change the
complexity of the rendering phase, which remains $O(n_s)$ -- although with a
smaller constant factor since computations are done per vertex instead of per
pixel.

\subsection{Quantitative evaluation}

Our quantitative results show that the O'Neal model underestimates the measured
values, by a large factor (about 3; the RMSE is
49.5
\unskip). The relative
luminance and the chromaticity also differ more from the measurements than
Nishita93 (especially near the Sun). However, this is mostly due to the number
of samples used for the numerical integration. With only 2 samples, as
originally proposed by O'Neal, the RMSE is $57.7$. With 32 samples, we get
almost the same results as Nishita93 (but this might not be an option for
real-time applications, especially on low-end GPUs).

\section{Haber Model}\label{sec_haber}

\newcommand{\relativeerrorcell}[2]{%
\begin{overpic}[width=0.2\mytablewidth]{relative_error_#2_#1}
\put(5,5){\footnotesize\input{program/output/figures/error_#2_#1.txt}}
\end{overpic}}

\newcommand{\relativeerrorrow}[1]{
\rotatebox{90}{\footnotesize #1 / 
\input{program/output/figures/sza_#1.txt}\unskip\degree} &
\begin{overpic}[width=0.2\mytablewidth]{relative_error_#1_nishita93}
\put(5,5){\footnotesize\input{program/output/figures/error_#1_nishita93.txt}}
\end{overpic} &
\begin{overpic}[width=0.2\mytablewidth]{relative_error_#1_nishita96}
\put(5,5){\footnotesize\input{program/output/figures/error_#1_nishita96.txt}}
\end{overpic} &
\begin{overpic}[width=0.2\mytablewidth]{relative_error_#1_preetham}
\put(5,5){\footnotesize\input{program/output/figures/error_#1_preetham.txt}}
\end{overpic} &
\begin{overpic}[width=0.2\mytablewidth]{relative_error_#1_oneal}
\put(5,5){\footnotesize\input{program/output/figures/error_#1_oneal.txt}}
\end{overpic} &
\begin{overpic}[width=0.2\mytablewidth]{relative_error_#1_haber}
\put(5,5){\footnotesize\input{program/output/figures/error_#1_haber.txt}}
\end{overpic} &
\begin{overpic}[width=0.2\mytablewidth]{relative_error_#1_bruneton}
\put(5,5){\footnotesize\input{program/output/figures/error_#1_bruneton.txt}}
\end{overpic} &
\begin{overpic}[width=0.2\mytablewidth]{relative_error_#1_elek}
\put(5,5){\footnotesize\input{program/output/figures/error_#1_elek.txt}}
\end{overpic} &
\begin{overpic}[width=0.2\mytablewidth]{relative_error_#1_hosek}
\put(5,5){\footnotesize\input{program/output/figures/error_#1_hosek.txt}}
\end{overpic} &
\begin{overpic}[width=0.2\mytablewidth]{relative_error_#1_libradtran}
\put(5,5){\footnotesize\input{program/output/figures/error_#1_libradtran.txt}}
\end{overpic}}

\newcommand{\relativeerrortable}{
\begin{mytabular}{p{0.04\mytablewidth}*{9}{p{0.198\mytablewidth}}}
&
\footnotesize\bfseries Nishita93 &
\footnotesize\bfseries Nishita96 &
\footnotesize\bfseries Preetham &
\footnotesize\bfseries O'Neal &
\footnotesize\bfseries Haber &
\footnotesize\bfseries Bruneton &
\footnotesize\bfseries Elek &
\footnotesize\bfseries Hosek &
\footnotesize\bfseries\leavevmode\color{blue} libRadtran\vspace{1mm}\\
\supplemental{
\rotatebox{90}{\footnotesize 09h30 / 
\input{program/output/figures/sza_09h30.txt}\unskip\degree} &
\begin{overpic}[width=0.2\mytablewidth]{relative_error_09h30_nishita93}
\put(5,5){\footnotesize\input{program/output/figures/error_09h30_nishita93.txt}}
\end{overpic} &
\begin{overpic}[width=0.2\mytablewidth]{relative_error_09h30_nishita96}
\put(5,5){\footnotesize\input{program/output/figures/error_09h30_nishita96.txt}}
\end{overpic} &
\begin{overpic}[width=0.2\mytablewidth]{relative_error_09h30_preetham}
\put(5,5){\footnotesize\input{program/output/figures/error_09h30_preetham.txt}}
\end{overpic} &
\begin{overpic}[width=0.2\mytablewidth]{relative_error_09h30_oneal}
\put(5,5){\footnotesize\input{program/output/figures/error_09h30_oneal.txt}}
\end{overpic} &
\begin{overpic}[width=0.2\mytablewidth]{relative_error_09h30_haber}
\put(5,5){\footnotesize\input{program/output/figures/error_09h30_haber.txt}}
\end{overpic} &
\begin{overpic}[width=0.2\mytablewidth]{relative_error_09h30_bruneton}
\put(5,5){\footnotesize\input{program/output/figures/error_09h30_bruneton.txt}}
\end{overpic} &
\begin{overpic}[width=0.2\mytablewidth]{relative_error_09h30_elek}
\put(5,5){\footnotesize\input{program/output/figures/error_09h30_elek.txt}}
\end{overpic} &
\begin{overpic}[width=0.2\mytablewidth]{relative_error_09h30_hosek}
\put(5,5){\footnotesize\input{program/output/figures/error_09h30_hosek.txt}}
\end{overpic} &
\begin{overpic}[width=0.2\mytablewidth]{relative_error_09h30_libradtran}
\put(5,5){\footnotesize\input{program/output/figures/error_09h30_libradtran.txt}}
\end{overpic}\\}
\supplemental{
\rotatebox{90}{\footnotesize 09h45 / 
\input{program/output/figures/sza_09h45.txt}\unskip\degree} &
\begin{overpic}[width=0.2\mytablewidth]{relative_error_09h45_nishita93}
\put(5,5){\footnotesize\input{program/output/figures/error_09h45_nishita93.txt}}
\end{overpic} &
\begin{overpic}[width=0.2\mytablewidth]{relative_error_09h45_nishita96}
\put(5,5){\footnotesize\input{program/output/figures/error_09h45_nishita96.txt}}
\end{overpic} &
\begin{overpic}[width=0.2\mytablewidth]{relative_error_09h45_preetham}
\put(5,5){\footnotesize\input{program/output/figures/error_09h45_preetham.txt}}
\end{overpic} &
\begin{overpic}[width=0.2\mytablewidth]{relative_error_09h45_oneal}
\put(5,5){\footnotesize\input{program/output/figures/error_09h45_oneal.txt}}
\end{overpic} &
\begin{overpic}[width=0.2\mytablewidth]{relative_error_09h45_haber}
\put(5,5){\footnotesize\input{program/output/figures/error_09h45_haber.txt}}
\end{overpic} &
\begin{overpic}[width=0.2\mytablewidth]{relative_error_09h45_bruneton}
\put(5,5){\footnotesize\input{program/output/figures/error_09h45_bruneton.txt}}
\end{overpic} &
\begin{overpic}[width=0.2\mytablewidth]{relative_error_09h45_elek}
\put(5,5){\footnotesize\input{program/output/figures/error_09h45_elek.txt}}
\end{overpic} &
\begin{overpic}[width=0.2\mytablewidth]{relative_error_09h45_hosek}
\put(5,5){\footnotesize\input{program/output/figures/error_09h45_hosek.txt}}
\end{overpic} &
\begin{overpic}[width=0.2\mytablewidth]{relative_error_09h45_libradtran}
\put(5,5){\footnotesize\input{program/output/figures/error_09h45_libradtran.txt}}
\end{overpic}\\}
\supplemental{
\rotatebox{90}{\footnotesize 10h00 / 
\input{program/output/figures/sza_10h00.txt}\unskip\degree} &
\begin{overpic}[width=0.2\mytablewidth]{relative_error_10h00_nishita93}
\put(5,5){\footnotesize\input{program/output/figures/error_10h00_nishita93.txt}}
\end{overpic} &
\begin{overpic}[width=0.2\mytablewidth]{relative_error_10h00_nishita96}
\put(5,5){\footnotesize\input{program/output/figures/error_10h00_nishita96.txt}}
\end{overpic} &
\begin{overpic}[width=0.2\mytablewidth]{relative_error_10h00_preetham}
\put(5,5){\footnotesize\input{program/output/figures/error_10h00_preetham.txt}}
\end{overpic} &
\begin{overpic}[width=0.2\mytablewidth]{relative_error_10h00_oneal}
\put(5,5){\footnotesize\input{program/output/figures/error_10h00_oneal.txt}}
\end{overpic} &
\begin{overpic}[width=0.2\mytablewidth]{relative_error_10h00_haber}
\put(5,5){\footnotesize\input{program/output/figures/error_10h00_haber.txt}}
\end{overpic} &
\begin{overpic}[width=0.2\mytablewidth]{relative_error_10h00_bruneton}
\put(5,5){\footnotesize\input{program/output/figures/error_10h00_bruneton.txt}}
\end{overpic} &
\begin{overpic}[width=0.2\mytablewidth]{relative_error_10h00_elek}
\put(5,5){\footnotesize\input{program/output/figures/error_10h00_elek.txt}}
\end{overpic} &
\begin{overpic}[width=0.2\mytablewidth]{relative_error_10h00_hosek}
\put(5,5){\footnotesize\input{program/output/figures/error_10h00_hosek.txt}}
\end{overpic} &
\begin{overpic}[width=0.2\mytablewidth]{relative_error_10h00_libradtran}
\put(5,5){\footnotesize\input{program/output/figures/error_10h00_libradtran.txt}}
\end{overpic}\\}

\rotatebox{90}{\footnotesize 10h15 / 
41
\unskip\degree} &
\begin{overpic}[width=0.2\mytablewidth]{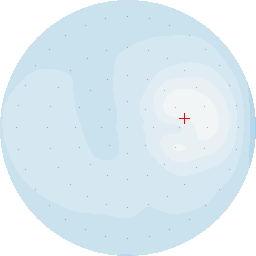}
\put(5,5){\footnotesize 24.7
}
\end{overpic} &
\begin{overpic}[width=0.2\mytablewidth]{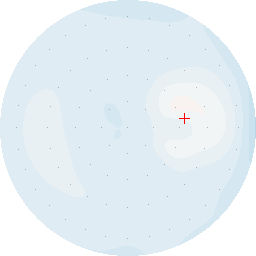}
\put(5,5){\footnotesize 17.7
}
\end{overpic} &
\begin{overpic}[width=0.2\mytablewidth]{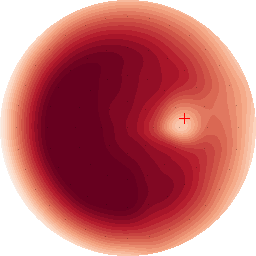}
\put(5,5){\footnotesize 92.4
}
\end{overpic} &
\begin{overpic}[width=0.2\mytablewidth]{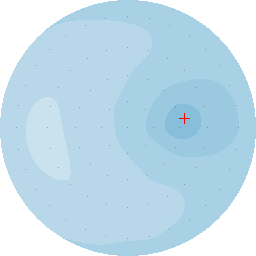}
\put(5,5){\footnotesize 49.7
}
\end{overpic} &
\begin{overpic}[width=0.2\mytablewidth]{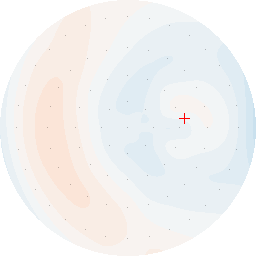}
\put(5,5){\footnotesize 12.6
}
\end{overpic} &
\begin{overpic}[width=0.2\mytablewidth]{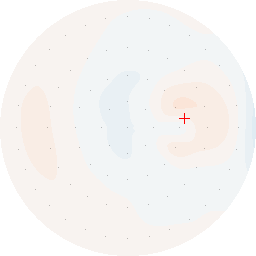}
\put(5,5){\footnotesize 9
}
\end{overpic} &
\begin{overpic}[width=0.2\mytablewidth]{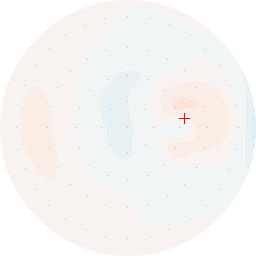}
\put(5,5){\footnotesize 9
}
\end{overpic} &
\begin{overpic}[width=0.2\mytablewidth]{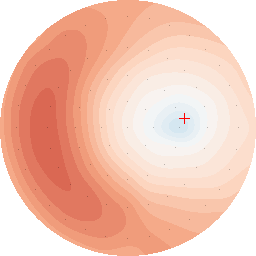}
\put(5,5){\footnotesize 39.9
}
\end{overpic} &
\begin{overpic}[width=0.2\mytablewidth]{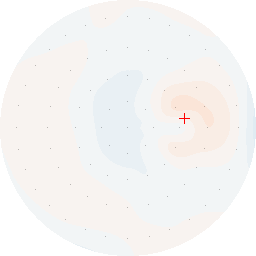}
\put(5,5){\footnotesize 9.3
}
\end{overpic}\\
\supplemental{
\rotatebox{90}{\footnotesize 10h30 / 
\input{program/output/figures/sza_10h30.txt}\unskip\degree} &
\begin{overpic}[width=0.2\mytablewidth]{relative_error_10h30_nishita93}
\put(5,5){\footnotesize\input{program/output/figures/error_10h30_nishita93.txt}}
\end{overpic} &
\begin{overpic}[width=0.2\mytablewidth]{relative_error_10h30_nishita96}
\put(5,5){\footnotesize\input{program/output/figures/error_10h30_nishita96.txt}}
\end{overpic} &
\begin{overpic}[width=0.2\mytablewidth]{relative_error_10h30_preetham}
\put(5,5){\footnotesize\input{program/output/figures/error_10h30_preetham.txt}}
\end{overpic} &
\begin{overpic}[width=0.2\mytablewidth]{relative_error_10h30_oneal}
\put(5,5){\footnotesize\input{program/output/figures/error_10h30_oneal.txt}}
\end{overpic} &
\begin{overpic}[width=0.2\mytablewidth]{relative_error_10h30_haber}
\put(5,5){\footnotesize\input{program/output/figures/error_10h30_haber.txt}}
\end{overpic} &
\begin{overpic}[width=0.2\mytablewidth]{relative_error_10h30_bruneton}
\put(5,5){\footnotesize\input{program/output/figures/error_10h30_bruneton.txt}}
\end{overpic} &
\begin{overpic}[width=0.2\mytablewidth]{relative_error_10h30_elek}
\put(5,5){\footnotesize\input{program/output/figures/error_10h30_elek.txt}}
\end{overpic} &
\begin{overpic}[width=0.2\mytablewidth]{relative_error_10h30_hosek}
\put(5,5){\footnotesize\input{program/output/figures/error_10h30_hosek.txt}}
\end{overpic} &
\begin{overpic}[width=0.2\mytablewidth]{relative_error_10h30_libradtran}
\put(5,5){\footnotesize\input{program/output/figures/error_10h30_libradtran.txt}}
\end{overpic}\\}
\supplemental{
\rotatebox{90}{\footnotesize 10h45 / 
\input{program/output/figures/sza_10h45.txt}\unskip\degree} &
\begin{overpic}[width=0.2\mytablewidth]{relative_error_10h45_nishita93}
\put(5,5){\footnotesize\input{program/output/figures/error_10h45_nishita93.txt}}
\end{overpic} &
\begin{overpic}[width=0.2\mytablewidth]{relative_error_10h45_nishita96}
\put(5,5){\footnotesize\input{program/output/figures/error_10h45_nishita96.txt}}
\end{overpic} &
\begin{overpic}[width=0.2\mytablewidth]{relative_error_10h45_preetham}
\put(5,5){\footnotesize\input{program/output/figures/error_10h45_preetham.txt}}
\end{overpic} &
\begin{overpic}[width=0.2\mytablewidth]{relative_error_10h45_oneal}
\put(5,5){\footnotesize\input{program/output/figures/error_10h45_oneal.txt}}
\end{overpic} &
\begin{overpic}[width=0.2\mytablewidth]{relative_error_10h45_haber}
\put(5,5){\footnotesize\input{program/output/figures/error_10h45_haber.txt}}
\end{overpic} &
\begin{overpic}[width=0.2\mytablewidth]{relative_error_10h45_bruneton}
\put(5,5){\footnotesize\input{program/output/figures/error_10h45_bruneton.txt}}
\end{overpic} &
\begin{overpic}[width=0.2\mytablewidth]{relative_error_10h45_elek}
\put(5,5){\footnotesize\input{program/output/figures/error_10h45_elek.txt}}
\end{overpic} &
\begin{overpic}[width=0.2\mytablewidth]{relative_error_10h45_hosek}
\put(5,5){\footnotesize\input{program/output/figures/error_10h45_hosek.txt}}
\end{overpic} &
\begin{overpic}[width=0.2\mytablewidth]{relative_error_10h45_libradtran}
\put(5,5){\footnotesize\input{program/output/figures/error_10h45_libradtran.txt}}
\end{overpic}\\}
\supplemental{
\rotatebox{90}{\footnotesize 11h00 / 
\input{program/output/figures/sza_11h00.txt}\unskip\degree} &
\begin{overpic}[width=0.2\mytablewidth]{relative_error_11h00_nishita93}
\put(5,5){\footnotesize\input{program/output/figures/error_11h00_nishita93.txt}}
\end{overpic} &
\begin{overpic}[width=0.2\mytablewidth]{relative_error_11h00_nishita96}
\put(5,5){\footnotesize\input{program/output/figures/error_11h00_nishita96.txt}}
\end{overpic} &
\begin{overpic}[width=0.2\mytablewidth]{relative_error_11h00_preetham}
\put(5,5){\footnotesize\input{program/output/figures/error_11h00_preetham.txt}}
\end{overpic} &
\begin{overpic}[width=0.2\mytablewidth]{relative_error_11h00_oneal}
\put(5,5){\footnotesize\input{program/output/figures/error_11h00_oneal.txt}}
\end{overpic} &
\begin{overpic}[width=0.2\mytablewidth]{relative_error_11h00_haber}
\put(5,5){\footnotesize\input{program/output/figures/error_11h00_haber.txt}}
\end{overpic} &
\begin{overpic}[width=0.2\mytablewidth]{relative_error_11h00_bruneton}
\put(5,5){\footnotesize\input{program/output/figures/error_11h00_bruneton.txt}}
\end{overpic} &
\begin{overpic}[width=0.2\mytablewidth]{relative_error_11h00_elek}
\put(5,5){\footnotesize\input{program/output/figures/error_11h00_elek.txt}}
\end{overpic} &
\begin{overpic}[width=0.2\mytablewidth]{relative_error_11h00_hosek}
\put(5,5){\footnotesize\input{program/output/figures/error_11h00_hosek.txt}}
\end{overpic} &
\begin{overpic}[width=0.2\mytablewidth]{relative_error_11h00_libradtran}
\put(5,5){\footnotesize\input{program/output/figures/error_11h00_libradtran.txt}}
\end{overpic}\\}

\rotatebox{90}{\footnotesize 11h15 / 
31
\unskip\degree} &
\begin{overpic}[width=0.2\mytablewidth]{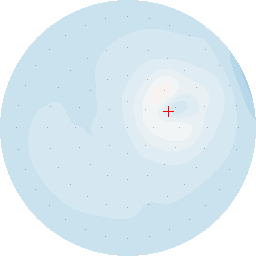}
\put(5,5){\footnotesize 26.8
}
\end{overpic} &
\begin{overpic}[width=0.2\mytablewidth]{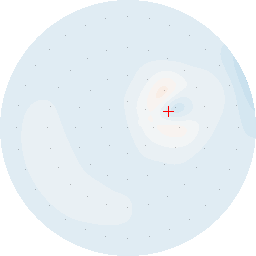}
\put(5,5){\footnotesize 18.7
}
\end{overpic} &
\begin{overpic}[width=0.2\mytablewidth]{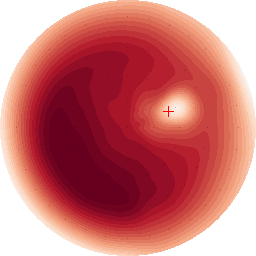}
\put(5,5){\footnotesize 88.4
}
\end{overpic} &
\begin{overpic}[width=0.2\mytablewidth]{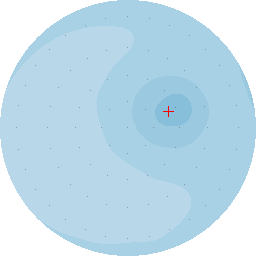}
\put(5,5){\footnotesize 49.8
}
\end{overpic} &
\begin{overpic}[width=0.2\mytablewidth]{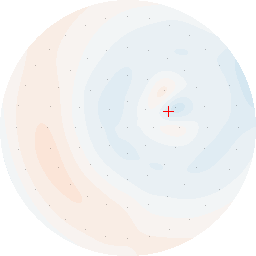}
\put(5,5){\footnotesize 15.2
}
\end{overpic} &
\begin{overpic}[width=0.2\mytablewidth]{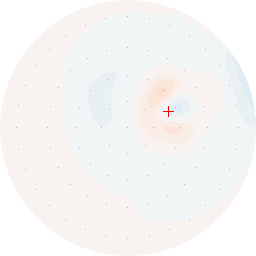}
\put(5,5){\footnotesize 11.1
}
\end{overpic} &
\begin{overpic}[width=0.2\mytablewidth]{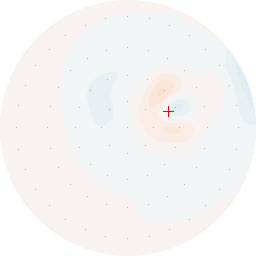}
\put(5,5){\footnotesize 11.1
}
\end{overpic} &
\begin{overpic}[width=0.2\mytablewidth]{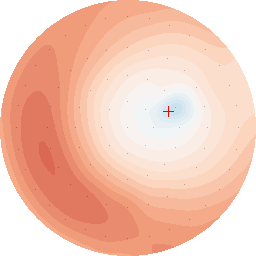}
\put(5,5){\footnotesize 42.8
}
\end{overpic} &
\begin{overpic}[width=0.2\mytablewidth]{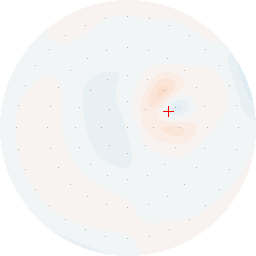}
\put(5,5){\footnotesize 11.1
}
\end{overpic}\\
\supplemental{
\rotatebox{90}{\footnotesize 11h30 / 
\input{program/output/figures/sza_11h30.txt}\unskip\degree} &
\begin{overpic}[width=0.2\mytablewidth]{relative_error_11h30_nishita93}
\put(5,5){\footnotesize\input{program/output/figures/error_11h30_nishita93.txt}}
\end{overpic} &
\begin{overpic}[width=0.2\mytablewidth]{relative_error_11h30_nishita96}
\put(5,5){\footnotesize\input{program/output/figures/error_11h30_nishita96.txt}}
\end{overpic} &
\begin{overpic}[width=0.2\mytablewidth]{relative_error_11h30_preetham}
\put(5,5){\footnotesize\input{program/output/figures/error_11h30_preetham.txt}}
\end{overpic} &
\begin{overpic}[width=0.2\mytablewidth]{relative_error_11h30_oneal}
\put(5,5){\footnotesize\input{program/output/figures/error_11h30_oneal.txt}}
\end{overpic} &
\begin{overpic}[width=0.2\mytablewidth]{relative_error_11h30_haber}
\put(5,5){\footnotesize\input{program/output/figures/error_11h30_haber.txt}}
\end{overpic} &
\begin{overpic}[width=0.2\mytablewidth]{relative_error_11h30_bruneton}
\put(5,5){\footnotesize\input{program/output/figures/error_11h30_bruneton.txt}}
\end{overpic} &
\begin{overpic}[width=0.2\mytablewidth]{relative_error_11h30_elek}
\put(5,5){\footnotesize\input{program/output/figures/error_11h30_elek.txt}}
\end{overpic} &
\begin{overpic}[width=0.2\mytablewidth]{relative_error_11h30_hosek}
\put(5,5){\footnotesize\input{program/output/figures/error_11h30_hosek.txt}}
\end{overpic} &
\begin{overpic}[width=0.2\mytablewidth]{relative_error_11h30_libradtran}
\put(5,5){\footnotesize\input{program/output/figures/error_11h30_libradtran.txt}}
\end{overpic}\\}
\supplemental{
\rotatebox{90}{\footnotesize 11h45 / 
27
\unskip\degree} &
\begin{overpic}[width=0.2\mytablewidth]{relative_error_11h45_nishita93}
\put(5,5){\footnotesize\input{program/output/figures/error_11h45_nishita93.txt}}
\end{overpic} &
\begin{overpic}[width=0.2\mytablewidth]{relative_error_11h45_nishita96}
\put(5,5){\footnotesize\input{program/output/figures/error_11h45_nishita96.txt}}
\end{overpic} &
\begin{overpic}[width=0.2\mytablewidth]{relative_error_11h45_preetham}
\put(5,5){\footnotesize\input{program/output/figures/error_11h45_preetham.txt}}
\end{overpic} &
\begin{overpic}[width=0.2\mytablewidth]{relative_error_11h45_oneal}
\put(5,5){\footnotesize\input{program/output/figures/error_11h45_oneal.txt}}
\end{overpic} &
\begin{overpic}[width=0.2\mytablewidth]{relative_error_11h45_haber}
\put(5,5){\footnotesize\input{program/output/figures/error_11h45_haber.txt}}
\end{overpic} &
\begin{overpic}[width=0.2\mytablewidth]{relative_error_11h45_bruneton}
\put(5,5){\footnotesize\input{program/output/figures/error_11h45_bruneton.txt}}
\end{overpic} &
\begin{overpic}[width=0.2\mytablewidth]{relative_error_11h45_elek}
\put(5,5){\footnotesize\input{program/output/figures/error_11h45_elek.txt}}
\end{overpic} &
\begin{overpic}[width=0.2\mytablewidth]{relative_error_11h45_hosek}
\put(5,5){\footnotesize\input{program/output/figures/error_11h45_hosek.txt}}
\end{overpic} &
\begin{overpic}[width=0.2\mytablewidth]{relative_error_11h45_libradtran}
\put(5,5){\footnotesize\input{program/output/figures/error_11h45_libradtran.txt}}
\end{overpic}\\}
\supplemental{
\rotatebox{90}{\footnotesize 12h00 / 
\input{program/output/figures/sza_12h00.txt}\unskip\degree} &
\begin{overpic}[width=0.2\mytablewidth]{relative_error_12h00_nishita93}
\put(5,5){\footnotesize\input{program/output/figures/error_12h00_nishita93.txt}}
\end{overpic} &
\begin{overpic}[width=0.2\mytablewidth]{relative_error_12h00_nishita96}
\put(5,5){\footnotesize\input{program/output/figures/error_12h00_nishita96.txt}}
\end{overpic} &
\begin{overpic}[width=0.2\mytablewidth]{relative_error_12h00_preetham}
\put(5,5){\footnotesize\input{program/output/figures/error_12h00_preetham.txt}}
\end{overpic} &
\begin{overpic}[width=0.2\mytablewidth]{relative_error_12h00_oneal}
\put(5,5){\footnotesize\input{program/output/figures/error_12h00_oneal.txt}}
\end{overpic} &
\begin{overpic}[width=0.2\mytablewidth]{relative_error_12h00_haber}
\put(5,5){\footnotesize\input{program/output/figures/error_12h00_haber.txt}}
\end{overpic} &
\begin{overpic}[width=0.2\mytablewidth]{relative_error_12h00_bruneton}
\put(5,5){\footnotesize\input{program/output/figures/error_12h00_bruneton.txt}}
\end{overpic} &
\begin{overpic}[width=0.2\mytablewidth]{relative_error_12h00_elek}
\put(5,5){\footnotesize\input{program/output/figures/error_12h00_elek.txt}}
\end{overpic} &
\begin{overpic}[width=0.2\mytablewidth]{relative_error_12h00_hosek}
\put(5,5){\footnotesize\input{program/output/figures/error_12h00_hosek.txt}}
\end{overpic} &
\begin{overpic}[width=0.2\mytablewidth]{relative_error_12h00_libradtran}
\put(5,5){\footnotesize\input{program/output/figures/error_12h00_libradtran.txt}}
\end{overpic}\\}
\supplemental{
\rotatebox{90}{\footnotesize 12h15 / 
23
\unskip\degree} &
\begin{overpic}[width=0.2\mytablewidth]{relative_error_12h15_nishita93}
\put(5,5){\footnotesize\input{program/output/figures/error_12h15_nishita93.txt}}
\end{overpic} &
\begin{overpic}[width=0.2\mytablewidth]{relative_error_12h15_nishita96}
\put(5,5){\footnotesize\input{program/output/figures/error_12h15_nishita96.txt}}
\end{overpic} &
\begin{overpic}[width=0.2\mytablewidth]{relative_error_12h15_preetham}
\put(5,5){\footnotesize\input{program/output/figures/error_12h15_preetham.txt}}
\end{overpic} &
\begin{overpic}[width=0.2\mytablewidth]{relative_error_12h15_oneal}
\put(5,5){\footnotesize\input{program/output/figures/error_12h15_oneal.txt}}
\end{overpic} &
\begin{overpic}[width=0.2\mytablewidth]{relative_error_12h15_haber}
\put(5,5){\footnotesize\input{program/output/figures/error_12h15_haber.txt}}
\end{overpic} &
\begin{overpic}[width=0.2\mytablewidth]{relative_error_12h15_bruneton}
\put(5,5){\footnotesize\input{program/output/figures/error_12h15_bruneton.txt}}
\end{overpic} &
\begin{overpic}[width=0.2\mytablewidth]{relative_error_12h15_elek}
\put(5,5){\footnotesize\input{program/output/figures/error_12h15_elek.txt}}
\end{overpic} &
\begin{overpic}[width=0.2\mytablewidth]{relative_error_12h15_hosek}
\put(5,5){\footnotesize\input{program/output/figures/error_12h15_hosek.txt}}
\end{overpic} &
\begin{overpic}[width=0.2\mytablewidth]{relative_error_12h15_libradtran}
\put(5,5){\footnotesize\input{program/output/figures/error_12h15_libradtran.txt}}
\end{overpic}\\}
\supplemental{
\rotatebox{90}{\footnotesize 12h30 / 
\input{program/output/figures/sza_12h30.txt}\unskip\degree} &
\begin{overpic}[width=0.2\mytablewidth]{relative_error_12h30_nishita93}
\put(5,5){\footnotesize\input{program/output/figures/error_12h30_nishita93.txt}}
\end{overpic} &
\begin{overpic}[width=0.2\mytablewidth]{relative_error_12h30_nishita96}
\put(5,5){\footnotesize\input{program/output/figures/error_12h30_nishita96.txt}}
\end{overpic} &
\begin{overpic}[width=0.2\mytablewidth]{relative_error_12h30_preetham}
\put(5,5){\footnotesize\input{program/output/figures/error_12h30_preetham.txt}}
\end{overpic} &
\begin{overpic}[width=0.2\mytablewidth]{relative_error_12h30_oneal}
\put(5,5){\footnotesize\input{program/output/figures/error_12h30_oneal.txt}}
\end{overpic} &
\begin{overpic}[width=0.2\mytablewidth]{relative_error_12h30_haber}
\put(5,5){\footnotesize\input{program/output/figures/error_12h30_haber.txt}}
\end{overpic} &
\begin{overpic}[width=0.2\mytablewidth]{relative_error_12h30_bruneton}
\put(5,5){\footnotesize\input{program/output/figures/error_12h30_bruneton.txt}}
\end{overpic} &
\begin{overpic}[width=0.2\mytablewidth]{relative_error_12h30_elek}
\put(5,5){\footnotesize\input{program/output/figures/error_12h30_elek.txt}}
\end{overpic} &
\begin{overpic}[width=0.2\mytablewidth]{relative_error_12h30_hosek}
\put(5,5){\footnotesize\input{program/output/figures/error_12h30_hosek.txt}}
\end{overpic} &
\begin{overpic}[width=0.2\mytablewidth]{relative_error_12h30_libradtran}
\put(5,5){\footnotesize\input{program/output/figures/error_12h30_libradtran.txt}}
\end{overpic}\\}
\supplemental{
\rotatebox{90}{\footnotesize 12h45 / 
\input{program/output/figures/sza_12h45.txt}\unskip\degree} &
\begin{overpic}[width=0.2\mytablewidth]{relative_error_12h45_nishita93}
\put(5,5){\footnotesize\input{program/output/figures/error_12h45_nishita93.txt}}
\end{overpic} &
\begin{overpic}[width=0.2\mytablewidth]{relative_error_12h45_nishita96}
\put(5,5){\footnotesize\input{program/output/figures/error_12h45_nishita96.txt}}
\end{overpic} &
\begin{overpic}[width=0.2\mytablewidth]{relative_error_12h45_preetham}
\put(5,5){\footnotesize\input{program/output/figures/error_12h45_preetham.txt}}
\end{overpic} &
\begin{overpic}[width=0.2\mytablewidth]{relative_error_12h45_oneal}
\put(5,5){\footnotesize\input{program/output/figures/error_12h45_oneal.txt}}
\end{overpic} &
\begin{overpic}[width=0.2\mytablewidth]{relative_error_12h45_haber}
\put(5,5){\footnotesize\input{program/output/figures/error_12h45_haber.txt}}
\end{overpic} &
\begin{overpic}[width=0.2\mytablewidth]{relative_error_12h45_bruneton}
\put(5,5){\footnotesize\input{program/output/figures/error_12h45_bruneton.txt}}
\end{overpic} &
\begin{overpic}[width=0.2\mytablewidth]{relative_error_12h45_elek}
\put(5,5){\footnotesize\input{program/output/figures/error_12h45_elek.txt}}
\end{overpic} &
\begin{overpic}[width=0.2\mytablewidth]{relative_error_12h45_hosek}
\put(5,5){\footnotesize\input{program/output/figures/error_12h45_hosek.txt}}
\end{overpic} &
\begin{overpic}[width=0.2\mytablewidth]{relative_error_12h45_libradtran}
\put(5,5){\footnotesize\input{program/output/figures/error_12h45_libradtran.txt}}
\end{overpic}\\}
\supplemental{
\rotatebox{90}{\footnotesize 13h00 / 
\input{program/output/figures/sza_13h00.txt}\unskip\degree} &
\begin{overpic}[width=0.2\mytablewidth]{relative_error_13h00_nishita93}
\put(5,5){\footnotesize\input{program/output/figures/error_13h00_nishita93.txt}}
\end{overpic} &
\begin{overpic}[width=0.2\mytablewidth]{relative_error_13h00_nishita96}
\put(5,5){\footnotesize\input{program/output/figures/error_13h00_nishita96.txt}}
\end{overpic} &
\begin{overpic}[width=0.2\mytablewidth]{relative_error_13h00_preetham}
\put(5,5){\footnotesize\input{program/output/figures/error_13h00_preetham.txt}}
\end{overpic} &
\begin{overpic}[width=0.2\mytablewidth]{relative_error_13h00_oneal}
\put(5,5){\footnotesize\input{program/output/figures/error_13h00_oneal.txt}}
\end{overpic} &
\begin{overpic}[width=0.2\mytablewidth]{relative_error_13h00_haber}
\put(5,5){\footnotesize\input{program/output/figures/error_13h00_haber.txt}}
\end{overpic} &
\begin{overpic}[width=0.2\mytablewidth]{relative_error_13h00_bruneton}
\put(5,5){\footnotesize\input{program/output/figures/error_13h00_bruneton.txt}}
\end{overpic} &
\begin{overpic}[width=0.2\mytablewidth]{relative_error_13h00_elek}
\put(5,5){\footnotesize\input{program/output/figures/error_13h00_elek.txt}}
\end{overpic} &
\begin{overpic}[width=0.2\mytablewidth]{relative_error_13h00_hosek}
\put(5,5){\footnotesize\input{program/output/figures/error_13h00_hosek.txt}}
\end{overpic} &
\begin{overpic}[width=0.2\mytablewidth]{relative_error_13h00_libradtran}
\put(5,5){\footnotesize\input{program/output/figures/error_13h00_libradtran.txt}}
\end{overpic}\\}

\rotatebox{90}{\footnotesize 13h15 / 
21
\unskip\degree} &
\begin{overpic}[width=0.2\mytablewidth]{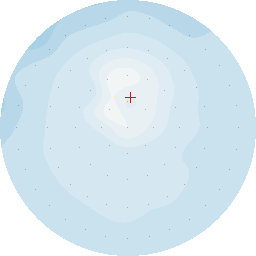}
\put(5,5){\footnotesize 28
}
\end{overpic} &
\begin{overpic}[width=0.2\mytablewidth]{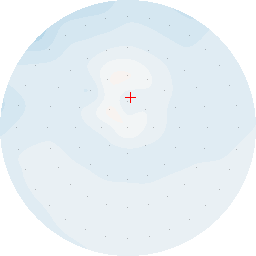}
\put(5,5){\footnotesize 17.4
}
\end{overpic} &
\begin{overpic}[width=0.2\mytablewidth]{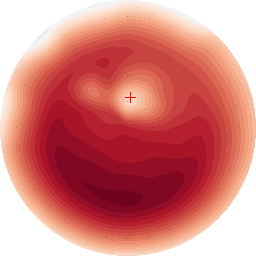}
\put(5,5){\footnotesize 83.3
}
\end{overpic} &
\begin{overpic}[width=0.2\mytablewidth]{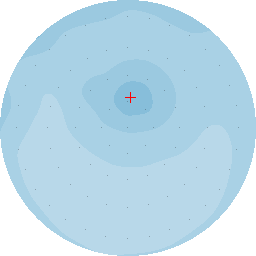}
\put(5,5){\footnotesize 48.4
}
\end{overpic} &
\begin{overpic}[width=0.2\mytablewidth]{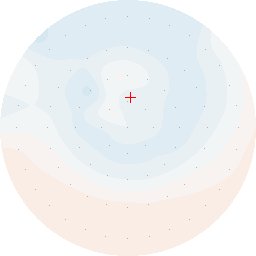}
\put(5,5){\footnotesize 14.9
}
\end{overpic} &
\begin{overpic}[width=0.2\mytablewidth]{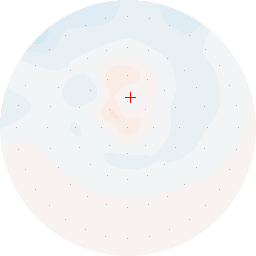}
\put(5,5){\footnotesize 11
}
\end{overpic} &
\begin{overpic}[width=0.2\mytablewidth]{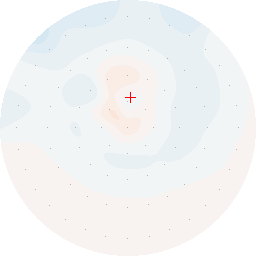}
\put(5,5){\footnotesize 11
}
\end{overpic} &
\begin{overpic}[width=0.2\mytablewidth]{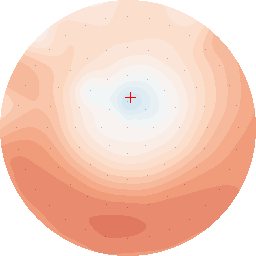}
\put(5,5){\footnotesize 40.5
}
\end{overpic} &
\begin{overpic}[width=0.2\mytablewidth]{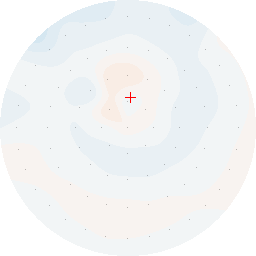}
\put(5,5){\footnotesize 10.9
}
\end{overpic}\supplemental{\\}%
\supplemental{
\rotatebox{90}{\footnotesize 13h30 / 
\input{program/output/figures/sza_13h30.txt}\unskip\degree} &
\begin{overpic}[width=0.2\mytablewidth]{relative_error_13h30_nishita93}
\put(5,5){\footnotesize\input{program/output/figures/error_13h30_nishita93.txt}}
\end{overpic} &
\begin{overpic}[width=0.2\mytablewidth]{relative_error_13h30_nishita96}
\put(5,5){\footnotesize\input{program/output/figures/error_13h30_nishita96.txt}}
\end{overpic} &
\begin{overpic}[width=0.2\mytablewidth]{relative_error_13h30_preetham}
\put(5,5){\footnotesize\input{program/output/figures/error_13h30_preetham.txt}}
\end{overpic} &
\begin{overpic}[width=0.2\mytablewidth]{relative_error_13h30_oneal}
\put(5,5){\footnotesize\input{program/output/figures/error_13h30_oneal.txt}}
\end{overpic} &
\begin{overpic}[width=0.2\mytablewidth]{relative_error_13h30_haber}
\put(5,5){\footnotesize\input{program/output/figures/error_13h30_haber.txt}}
\end{overpic} &
\begin{overpic}[width=0.2\mytablewidth]{relative_error_13h30_bruneton}
\put(5,5){\footnotesize\input{program/output/figures/error_13h30_bruneton.txt}}
\end{overpic} &
\begin{overpic}[width=0.2\mytablewidth]{relative_error_13h30_elek}
\put(5,5){\footnotesize\input{program/output/figures/error_13h30_elek.txt}}
\end{overpic} &
\begin{overpic}[width=0.2\mytablewidth]{relative_error_13h30_hosek}
\put(5,5){\footnotesize\input{program/output/figures/error_13h30_hosek.txt}}
\end{overpic} &
\begin{overpic}[width=0.2\mytablewidth]{relative_error_13h30_libradtran}
\put(5,5){\footnotesize\input{program/output/figures/error_13h30_libradtran.txt}}
\end{overpic}\\}
\end{mytabular}}

\myfigure{\relativeerrortable

\includegraphics[width=1.5\mytablewidth]{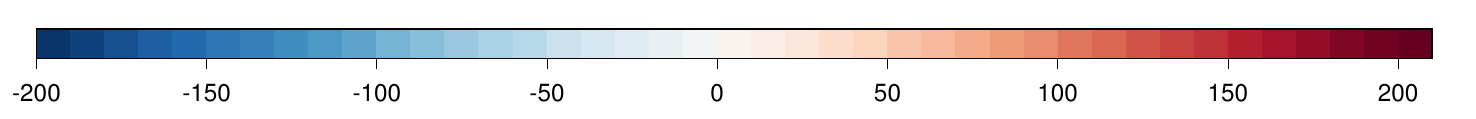}\vspace{-0.5cm}}{%
\caption{{\bf Relative error}. The relative error compared with the
measurements, in \% and using the same color scale as in~\cite{Kider14},
computed at the 81 sampling points (and summed equally over the common range
supported by all models, i.e. between $360$ and $720\,nm$), and then
interpolated with spherical bicubic interpolation. The bottom left number is the
RMSE in $mW/(m^2.sr.nm)$ (computed over the 81 sampling points and all the
wavelengths between $360$ and $720\,nm$).}\label{fig:relative_error}}

\subsection{Overview}

The Haber model\cite{Haber05} stores the atmospheric properties in $n_s$ layers
of constant optical depths, as in Nishita93, and uses a 3D grid to compute the
successive orders of multiple scattering, as in Nishita96. However, Haber et al.
use a $n_r\times n_{\theta}\times n_{\phi}$ grid based on spherical coordinates
centered on the viewer, instead of the $n_x\times n_y\times n_z$ Cartesian grid
used in\cite{Nishita96}. Also, they compute multiple scattering more precisely,
by integrating over all directions (in fact over all cells) at each grid cell
(instead of using only 8 coplanar specific directions as in\cite{Nishita96}).
The main approximations in their model are:
\begin{itemize}
\item the approximation of the Rayleigh phase function with the isotropic phase
function, for all scattering orders,
\item the approximation of the Mie phase function with a $g$ $\delta$-peak and a
$(1-g)$ isotropic lobe, except for single scattering.
\end{itemize}
The latter approximation reduces to the use of an effective Mie scattering
coefficient (see Eq. (2) in\cite{Haber05}), which must also be used in the Mie
extinction coefficient to avoid energy losses.

Once the 3D grid is computed, rendering a pixel is done with a numerical
integration along the corresponding view ray, with a lookup in the 3D grid per
integration sample (plus a computation of the transmittance between the viewer
and this sample, which requires a nested integral).

Haber et al. also take into account the refractive index of air, yielding curved
light paths, as well as ozone absorption. They also propose algorithmic
optimizations for the main loop over all cell pairs -- a naive implementation
would have a $O(n_r^2n_{\theta}^2n_{\phi}^2n_s)$ complexity.

\subsection{Our implementation}

In our implementation we neglect the ozone absorption and the refractive index
of air. We also ignore the ground albedo, as in\cite{Haber05}. We use $n_s=50$
layers and $n_{\phi}=72$ azimuthal directions for the 3D grid (in\cite{Haber05}
between 20-50 layers and 72-120 directions are used), yielding a total of 65268
cells, and 4 scattering orders. Finally, we do not use the effective Mie
scattering coefficient approximation. Instead, we approximate the Mie phase
function with the isotropic phase function and use the raw Mie scattering
coefficient (except for single scattering, where we use the Cornette-Shanks
phase function, as in\cite{Haber05}). We found this method less error prone
(it is hard to make sure the effective scattering coefficient is not used for
single scattering, especially in the final pass that combines the single and
multiple scattering contributions) and slightly more accurate.

\subsection{Qualitative evaluation}

The 3D grid of cells used in the Haber model is computed for a specific sun
zenith angle, and a specific viewer altitude. Therefore, this model is limited
to views on or near the ground (views from space show points with different
local sun zenith angles simultaneously). On the other hand, since a pixel is
rendered with a numerical integration along the view ray, it is easy to support
aerial perspective. Sun directions below the horizon are also supported. In
fact, this was the main motivation of this model.

The precomputation phase for the 3D grid, assuming a fixed number of scattering
orders, has $O(n_r^2n_{\theta}^2n_{\phi}^2)$ and $O(n_rn_{\theta}n_{\phi})$ time
and memory complexity, respectively, thanks to the algorithmic optimizations
described in\cite{Haber05}. Rendering a pixel, as described in\cite{Haber05},
has a $O(n_r n_s)$ time complexity (because the transmittance is not
precomputed, unlike in the previous models), but this can easily be reduced to
$O(n_r)$ with precomputation.

\subsection{Quantitative evaluation}

\begin{figure}
\begin{center}
\includegraphics[width=\columnwidth]{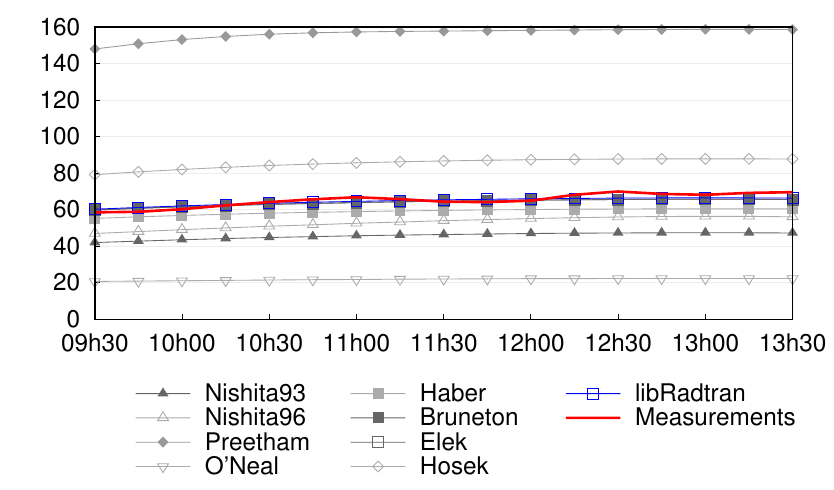}\vspace{-0.5cm}
\end{center}
\caption{{\bf Sky irradiance}. The sky irradiance in $W.m^{-2}$ as a function of
the time of day (spectral irradiance integrated over the 360-720 $nm$
range).}\label{fig:sky_irradiance}
\end{figure}

Our quantitative results, in particular in \cref{fig:sky_irradiance}, show that
the Haber model slightly underestimates the measured values, but less than the
Nishita96 model (the RMSE is
14.7
\vs
18.3
 for Nishita96). This
seems logical, since this model takes more scattering orders into account, and
uses less approximations to compute each order (but, on the other hand, ignores
the ground albedo). \cref{fig:relative_error} shows that this underestimation is
not uniform, and some areas show in fact an overestimation. This is best seen at
about 90 degrees from the sun, and is due to the approximation of the Rayleigh
phase function with the isotropic phase function (removing this approximation
for single scattering only is easy, does not require much more memory, and gives
significantly better results).

\section{Bruneton Model}\label{sec_bruneton}

\subsection{Overview}

The Bruneton model\cite{Bruneton08} precomputes the multiple scattering orders
in sequence, as in the Nishita96 and Haber models. However, instead of doing
this precomputation for a single sun zenith angle, Bruneton et al. do it for
$n_{\theta_s}$ sun zenith angles, including below the horizon. As a consequence,
they need a 4D grid instead of a 3D one as in\cite{Nishita96} and\cite{Haber05}.
While Nishita96 uses a Cartesian grid, and Haber a grid based on spherical
coordinates, the "grid" used in Bruneton et al. comes from the parameterization
used ($n_r$ altitudes, $n_{\theta}$ and $n_{\theta_s}$ view and sun zenith
angles and $n_{\gamma}$ view-sun angles), and has no simple geometric
representation. Bruneton et al. take the ground albedo into account (as
Nishita96 but unlike Haber), as well as the Rayleigh and Cornette-Shanks phase
functions (Haber et al. use some isotropic approximations to save memory and
computation time). As in\cite{Haber05}, multiple scattering is computed with an
integral over all directions at each view ray sample (whereas Nishita96
approximates this using only 8 coplanar directions).

\subsection{Our implementation}

Our implementation is directly based on the authors source code, simply extended
to work with $n_\lambda$ (\ie 40) dimensional vectors instead of RGB values.
Because of this, we didn't use the approximation proposed in Section 4
of\cite{Bruneton08}, which keeps only one spectral value for the single Mie
scattering in order to be able to pack the Rayleigh and Mie terms in a single
GPU vector. Instead, we store them separately in two 40-dimensional vectors.
Finally, we also adapted the code to run on CPU, to make use of the same
framework and double precision as in our implementation of the other models. We
used the same grid resolution and number of integration samples as in the
original code.

\subsection{Qualitative evaluation}

The Bruneton model supports all viewpoints from ground to space, aerial
perspective and sun directions below the horizon. 

The most costly precomputation step is the computation of $\Delta J$ in
Algorithm 4.1 of\cite{Bruneton08}. It involves an integral over all directions
(of $O(n_{\theta'}n_{\phi'})$ complexity) for all cells of the 4D $\Delta J$
table, yielding a $O(n_rn_{\theta}n_{\theta_s}n_{\gamma}n_{\theta'}n_{\phi'})$
and $O(n_rn_{\theta}n_{\theta_s}n_{\gamma})$ time and memory complexity,
respectively. At render time, the complexity is simply $O(1)$.

\subsection{Quantitative evaluation}

Our quantitative results show that the Bruneton model gives more accurate
results than the Haber model (the RMSE decreases from
14.7
 to
11.3
\unskip). The sky
radiance is sometimes underestimated (in particular very close to the sun, as
all the models), and sometimes overestimated (\eg near the horizon in the
opposite sun direction). However, the Bruneton model gives very close results to
the libRadtran reference model, in all quantitative evaluations (see
\quantitativefigs). This suggests that the difference with the measurements
comes from the input parameters used (such as the Mie phase function), and/or
from the neglected physical processes (molecular absorption, polarization, etc),
rather than from approximations used to solve the physical processes that are
taken into account (\eg insufficient number of samples in some numerical
integrals). In particular, the underestimation near the sun, observed in all
models including libRadtran, is almost certainly due to a missing strong forward
scattering peak in the Cornette-Shanks phase function used to approximate the
real Mie phase function.

\section{Elek Model}\label{sec_elek}

\subsection{Overview}

The Elek model\cite{Elek10} extends the Bruneton model to arbitrary atmospheres
and to oceans. It computes the multiple scattering orders in sequence, as in the
Nishita96, Haber and Bruneton models, using mostly the same algorithm, 4D tables
and texture parameterizations as in the Bruneton model. However, instead of
doing these precomputations on 3 wavelengths, Elek et al. perform them on 15
wavelengths, and convert the results to RGB at the very end of the
precomputation phase (leaving the sky rendering phase unchanged).

\subsection{Our Implementation}

When running all the models with 40 wavelengths, which is what we did for our
quantitative evaluations as explained in \cref{sec_quantitativeeval}, the
Bruneton and Elek models become identical. For this reason, we used the same
implementation for both models.

\subsection{Qualitative evaluation}

The Elek model supports the same viewpoints as the Bruneton model. It also has
the same algorithmic complexity, when both models are run spectrally.

\subsection{Quantitative evaluation}

Our quantitative results in \quantitativefigs\ are exactly the same as those of
the Bruneton model, since the Elek and Bruneton models are identical when run
spectrally. However, when using the original number of wavelengths proposed in
each model, the Elek model is more accurate (\cf \cref{sec_rgborspectral}).

\section{Hosek Model}\label{sec_hosek}

\subsection{Overview}

The Hosek model\cite{Hosek12} is an analytical model similar to the Preetham
model. The main differences are:
\begin{itemize}
\item an improved analytical formula, with a few more degrees of freedom for the
fitting phase,
\item a fully spectral model (instead of using CIE XYZ values converted at the
end to a radiance spectrum), allowing a user specified extraterrestrial solar
spectrum,
\item new model parameters, in addition to turbidity, for the spectral ground
albedo.
\end{itemize}
Also, the analytical functions were fitted to results obtained with a
path-tracer, \emph{a priori} more accurate than the Nishita96 model used by
Preetham.

\subsection{Our implementation}

Our implementation is a thin wrapper around the authors source code. As with the
Preetham model, we set the turbidity to $2.53$. However, unlike with the
Preetham model, we also set the extraterrestrial solar spectrum and the spectral
ground albedo to the values used in all other models, taking advantage of the
fact that the Hosek model handles these parameters.

\subsection{Qualitative evaluation}

The Hosek model is limited to views from the ground, because the analytical
functions were fitted for an observer at the ground level. It is also limited to
sun directions above the horizon (as explained in\cite{Hosek12}, supporting
these would require different analytic formulae, because the sky radiance
pattern is significantly different at sunrise / sunset than during the day).
Finally, unlike Preetham, Hosek et al. do not provide a separate model for the
aerial perspective, which is thus not supported.

The time and memory complexity to render a pixel is simply $O(1)$. There is no
precomputation phase at all, if one simply wants to use the model as is.
However, if one wants to change some atmospheric parameter, it is necessary to
recompute the sky radiance for many view directions and sun directions, and
perform a new non linear fitting.

\subsection{Quantitative evaluation}

Our quantitative results show that the Hosek model overestimates the measured
values, by a large factor (but not as large as for the Preetham model), except
near the sun where, on the contrary, it underestimates the measured values. The
RMSE is 41.5
\unskip.

This is surprising, since this model was fitted using results from a path-tracer
using the same physical equations as in all the other models. In any case, we
believe that the family of analytical functions used by Hosek et al., fitted
directly to the measurements, could produce much better results (in other words,
the observed discrepancy is probably not due to the choice of the family of
functions used for the fitting).

\section{Perceptual study}\label{sec_study}

\newcommand{\userstudyrow}[1]{
\includegraphics[width=0.25\mytablewidth]{sky_image_#1_nishita93}&
\includegraphics[width=0.25\mytablewidth]{sky_image_#1_nishita96}&
\includegraphics[width=0.25\mytablewidth]{sky_image_#1_preetham}&
\includegraphics[width=0.25\mytablewidth]{sky_image_#1_oneal}&
\includegraphics[width=0.25\mytablewidth]{sky_image_#1_haber}&
\includegraphics[width=0.25\mytablewidth]{sky_image_#1_bruneton}&
\includegraphics[width=0.25\mytablewidth]{sky_image_#1_elek}&
\includegraphics[width=0.25\mytablewidth]{sky_image_#1_hosek}}

\newcommand{\userstudy}{
\begin{mytabular}{*{8}{p{0.25\mytablewidth}}}
\footnotesize\bfseries Nishita93 &
\footnotesize\bfseries Nishita96 &
\footnotesize\bfseries Preetham &
\footnotesize\bfseries O'Neal &
\footnotesize\bfseries Haber &
\footnotesize\bfseries Bruneton &
\footnotesize\bfseries Elek &
\footnotesize\bfseries Hosek\vspace{1mm} \\
\userstudyrow{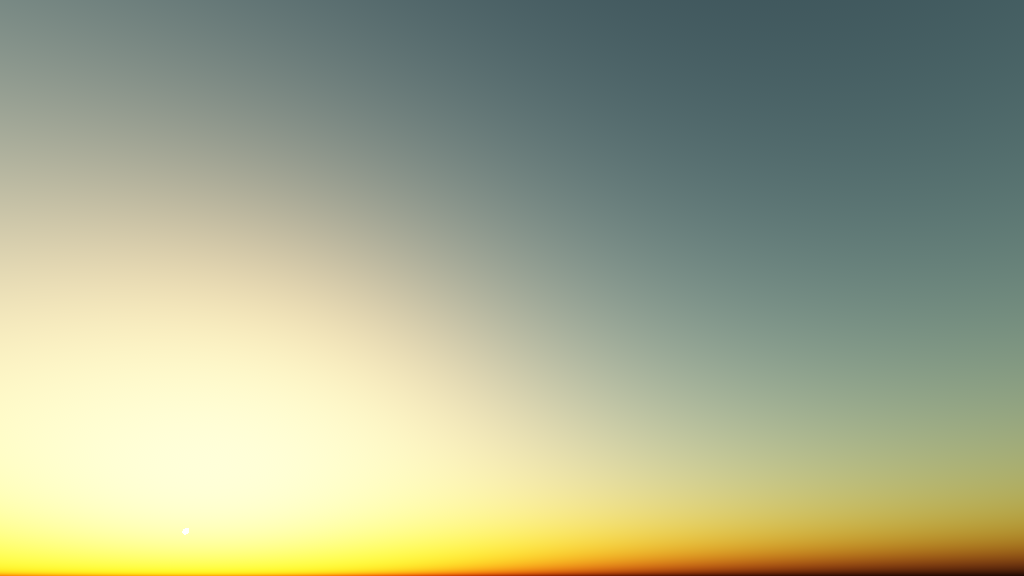}\\
\userstudyrow{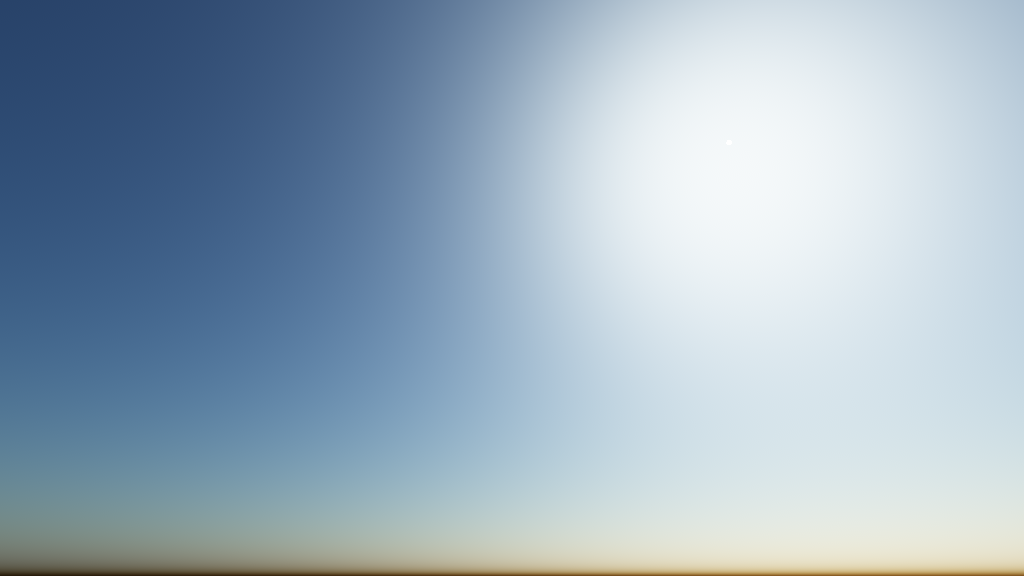}
\end{mytabular}}

\myfigure{\userstudy\vspace{-1mm}}{%
\caption{{\bf Perceptual study inputs}. The images that participants have been
asked to compare by pairs (within each row). Since the Nishita93 and Nishita96
images are almost identical, we excluded the Nishita93 model from the
study.\label{fig:userstudy}}}

While some applications require physical accuracy (\eg energy studies), other
applications only require the sky to look realistic to users. It is thus
interesting to evaluate how the models are perceived by users. For this we did a
perceptual study, which is presented here.

\subsection{Experimental procedure}

In order to find which models look the most realistic to users, we asked them to
compare pairs of images produced by different models but with identical
atmospheric parameters. We asked each participant of the study to compare all
the possible pairs, in random order. For each pair, the task was "Click on the
image that you find the most realistic".

Before running the experiment, we generated the images to compare as follows:
\begin{itemize}
\item we chose a viewport including the horizon and the sun (the two areas where
the models differ the most), but excluding the ground (not all the models
support aerial perspective, which is needed to render the ground in a realistic
way),
\item we used the same atmospheric parameters as in \cref{sec_reference} for all
the models and we ran each model with the original number of wavelengths
proposed by their authors (as described in \cref{sec_rgbrender}), in order to
evaluate the perceptual impact of this parameter,
\item we scaled the results with a uniform per-model factor, before
tone-mapping, so that each model gives the same sky irradiance (indeed, we are
not interested in absolute values here).
\end{itemize}
The resulting images are shown in \cref{fig:userstudy} (we used two scenes: a
sunrise and a morning sky). Since the images for Nishita93 and Nishita96 are
almost identical, we decided to exclude the Nishita93 model from the experiment.

Our experiment was conducted with 25 participants in a dark room, on a
calibrated HP ZR2440w monitor (24-inch, gamma $2.2$, color temperature
$6500\,K$, viewing distance $70\,\mathrm{cm}$). We also did the same experiment
online with 105 participants provided by Prolific.ac (using their desktop
computer in an uncontrolled environment -- see the supplemental material).

\subsection{Analysis methodology}

The result of the study is a preference matrix $A$ where $A_{i,j}$ is the number
of participants who find the model in row $i$ more realistic than the model in
column $j$. From this we can compute a score $a_i=\sum_{j\neq i} A_{i,j}$ for
each model, which we can use to rank the models. However, not all score
differences are statistically significant. We consider that there is a
statistically significant difference between the realism of two models if their
score difference is larger than the value $\lceil R_c \rceil$ defined in
\cite{Setyawan04} (following the same methodology as \cite{Setyawan04,Ledda05}).

\begin{table}
\caption{{\bf Perceptual study results}. The preference matrix, the model scores
(total number of votes, in bold) and the model groups, for the sunrise (top) and
morning (bottom) sky perceptual studies. $M_1,\ldots,M_7$ correspond
respectively to the Nishita96, Preetham, O'Neal, Haber, Bruneton, Elek and Hosek
models.\label{tab:prefmatrix}}
\begin{center}
\begin{tabular}{|c||c|c|c|c|c|c|c|c|}\hline
 & $M_1$ & $M_2$ & $M_3$ & $M_4$ & $M_5$ & $M_6$ & $M_7$ & {\bf Total}\\ \hline \hline
$M_1$ & - & 18 & 12 & 3 & 13 & 9 & 11 & {\bf 66}\\ \hline
$M_2$ & 7 & - & 12 & 1 & 4 & 3 & 4 & {\bf 31}\\ \hline
$M_3$ & 13 & 13 & - & 6 & 7 & 10 & 11 & {\bf 60}\\ \hline
$M_4$ & 22 & 24 & 19 & - & 20 & 17 & 20 & {\bf 122}\\ \hline
$M_5$ & 12 & 21 & 18 & 5 & - & 14 & 18 & {\bf 88}\\ \hline
$M_6$ & 16 & 22 & 15 & 8 & 11 & - & 14 & {\bf 86}\\ \hline
$M_7$ & 14 & 21 & 14 & 5 & 7 & 11 & - & {\bf 72}\\ \hline
\end{tabular}

\end{center}
\begin{center}
\nunderline{\hspace{2.05mm}$M_2$\hspace{2.05mm}}%
\nunderline[2]{\hspace{2.05mm}$M_3$\hspace{2.05mm}}%
\nunderline{\nunderline{\hspace{2.05mm}$M_1$\hspace{2.05mm}}}%
\nunderline{\nunderline{\hspace{2.05mm}$M_7$\hspace{2.05mm}}}%
\nunderline{\nunderline{\hspace{2.05mm}$M_6$\hspace{2.05mm}}}%
\nunderline{\hspace{2.05mm}$M_5$\hspace{2.05mm}}%
\nunderline[2]{\hspace{2.05mm}$M_4$\hspace{2.05mm}}%
\hspace{2.05mm} ($\lceil R_c \rceil=28$)

\end{center}
\begin{center}
\begin{tabular}{|c||c|c|c|c|c|c|c|c|}\hline
 & $M_1$ & $M_2$ & $M_3$ & $M_4$ & $M_5$ & $M_6$ & $M_7$ & {\bf Total}\\ \hline \hline
$M_1$ & - & 18 & 19 & 14 & 10 & 13 & 6 & {\bf 80}\\ \hline
$M_2$ & 7 & - & 14 & 6 & 4 & 5 & 5 & {\bf 41}\\ \hline
$M_3$ & 6 & 11 & - & 8 & 4 & 11 & 7 & {\bf 47}\\ \hline
$M_4$ & 11 & 19 & 17 & - & 8 & 13 & 10 & {\bf 78}\\ \hline
$M_5$ & 15 & 21 & 21 & 17 & - & 18 & 13 & {\bf 105}\\ \hline
$M_6$ & 12 & 20 & 14 & 12 & 7 & - & 5 & {\bf 70}\\ \hline
$M_7$ & 19 & 20 & 18 & 15 & 12 & 20 & - & {\bf 104}\\ \hline
\end{tabular}

\end{center}
\begin{center}
\nunderline{\hspace{2.05mm}$M_2$\hspace{2.05mm}}%
\nunderline{\nunderline{\hspace{2.05mm}$M_3$\hspace{2.05mm}}}%
\nunderline{\nunderline[2]{\hspace{2.05mm}$M_6$\hspace{2.05mm}}}%
\nunderline[2]{\nunderline{\hspace{2.05mm}$M_4$\hspace{2.05mm}}}%
\nunderline[2]{\nunderline{\hspace{2.05mm}$M_1$\hspace{2.05mm}}}%
\nunderline{\hspace{2.05mm}$M_7$\hspace{2.05mm}}%
\nunderline{\hspace{2.05mm}$M_5$\hspace{2.05mm}}%
\hspace{2.05mm} ($\lceil R_c \rceil=28$)

\end{center}
\end{table}

\subsection{Experiment results}

\begin{table*}
\renewcommand{\arraystretch}{1.3}
\caption{{\bf Qualitative comparison}. Summary of the qualitive evaluation of
the 8 clear sky models. The precomputation time and memory complexity for the
Haber and Nishita96 models is for a single Sun zenith angle $\theta_s$ only --
the Bruneton and Elek models precomputations are for $n$ such angles and the
Nishita93 precomputations are independent of $\theta_s$.}
\label{fig:qualitative_comparison}
\centering
\begin{tabular}{|r||c|c|c|c|c|c|c|c|}
\hline
\bfseries & \bfseries Supported & \bfseries Aerial & \bfseries Sunset &
\bfseries Scattering & \bfseries Precompute & \bfseries Precompute &
\bfseries Render & \bfseries RMSE \\
\bfseries Model & \bfseries viewpoints & \bfseries perspective & \bfseries
sunrise & \bfseries orders & \bfseries time & \bfseries memory & \bfseries time
& \bfseries $mW/(m^2.sr.nm)$ \\
\hline
Nishita93 & all & yes & yes & 1 & $O(n^3)$ & $O(n^2)$ & $O(n)$ &
26.6
 \\
\hline
Nishita96 & in atmosphere & yes & yes & 2 & $O(n^3)$ & $O(n^3)$ & $O(n)$ &
18.3
 \\
\hline
Preetham & ground only & yes & no & 2 & $0$ & $0$ & $O(1)$ &
88.1
 \\
\hline
O'Neal & all & yes & yes & $1$ & $0$ & $0$ & $O(n)$ &
49.5
 \\
\hline
Haber & ground only & yes & yes & all & $O(n^6)$ & $O(n^3)$ & $O(n^2)$ &
14.7
 \\
\hline
Bruneton & all & yes & yes & all & $O(n^6)$ & $O(n^4)$ & $O(1)$ &
11.3
 \\
\hline
Elek & all & yes & yes & all & $O(n^6)$ & $O(n^4)$ & $O(1)$ &
11.3
 \\
\hline
Hosek & ground only & no & no & all & $0$ & $0$ & $O(1)$ &
41.5
 \\
\hline
\end{tabular}
\end{table*}

The preference matrix and model scores of each scene are shown in
\cref{tab:prefmatrix}, together with the statistically significant score
difference $\lceil R_c \rceil$, for the significance level $\alpha=0.05$.
Grouping together the models whose score difference is less than this threshold
gives the groups shown in \cref{tab:prefmatrix}.

For the sunrise scene the participants find the Haber model significantly more
realistic than the Bruneton, Elek, Hosek and Nishita96 models (between which
there is no statistically significant differences), while the Preetham and
O'Neal models are perceived as the less realistic.

For the morning scene the participants find the Bruneton, Hosek, Nishita96 and
Haber models more realistic (with no statistically significant differences
between them), followed by the Elek model, with the Preetham and O'Neal models
perceived again as the less realistic.

The online experiment gives almost the same ranking as the laboratory experiment
for the morning scene, but gives more different results for the sunrise scene
(Preetham and O'Neal are still perceived as ones of the less realistic models,
but the Haber model is no longer perceived as significantly more realistic than
the Bruneton, Elek and Hosek models -- see the supplemental material). This
suggests that the viewing conditions play an important role in the perceived
realism of sunrise and sunset scenes.

Overall, these results show that the less physically accurate models, according
to our results in \cref{fig:qualitative_comparison}, \ie the Preetham and O'Neal
models, are also perceived as less realistic by the participants. Conversely,
the more physically accurate models are perceived as more realistic. However, it
seems that participants perceive models whose physical accuracy is "good enough"
as equally realistic.

\section{Discussion}\label{sec_discussion}

This section compares our results with previous work, summarizes the advantages
and drawbacks of the models presented above, and discusses some possible ways to
improve the accuracy of clear sky models in Computer Graphics.

\subsection{Comparison with previous work}

According to our results, the clear sky models, sorted in increasing order of
physical accuracy in $W/(m^2.sr.nm)$, are Preetham, O'Neal, Hosek, Nishita93,
Nishita96, Haber, Bruneton and Elek (\cf \cref{fig:qualitative_comparison}). We
also get a very similar order when measured after RGB conversion and tone
mapping (\cf \cref{fig:rgb_rendering}). According to Kider et al., the most
accurate models are Nishita93 and Nishita96 (\cf Figs 1, 12 and 15
in\cite{Kider14}). This is problematic, since both papers aim at comparing the
models with each other and with a ground thruth. We believe our results are more
plausible because:
\begin{itemize}
\item the order we get corresponds to a decreasing order of approximations: from
the models which approximate all the parameters with a single one, to the models
with more parameters and less and less approximations for multiple scattering.
In other words, we find that the less approximations are made the more accurate
is the result, which seems natural,
\item we get the same results for Preetham as in\cite{Zotti07} (compare
\cref{fig:relative_luminance} and \cref{fig:absolute_luminance} with Figs. 6 and
7 in\cite{Zotti07}),
\item we get similar results for the single and double scattering components of
the Nishita96, Haber and Bruneton models, although the 3 implementations are
completely different (see the supplemental material),
\item we get almost identical results for the Bruneton and Elek models as
libRadtran.
\end{itemize}

\subsection{Advantages and drawbacks}

The analytical models, \ie the Preetham and Hosek models, are fast, memory
efficient, easy to implement and don't require a precomputation phase (see
\cref{fig:qualitative_comparison}). On the other hand, they are limited to
ground level views and require a separate model for aerial perspective (which is
not always provided). They also have very few parameters, which limits the range
of atmospheric conditions they can reproduce. Our results confirm the results
from\cite{Zotti07,Kider14}, \ie that the Hosek model is more accurate than the
Preetham model.

Conversely, the numerical models, \ie the Nishita, O'Neal, Haber, Bruneton and
Elek models, are slower, use more memory, are more complex and generally require
a precomputation phase, but they can support more viewpoints, include aerial
perspective, and provide more physical parameters (and can thus better fit real
data). 

Amongst the numerical models, the Nishita93 and O'Neal models are the simplest
and the ones which require the less memory. Their main drawback is that they
ignore multiple scattering, which limits their accuracy, especially for sunrise
and sunset. The Nishita96 model adds support for multiple scattering, with some
coarse approximations. This increases its accuracy, at the price of increased
complexity, rendering times and memory needs. Another drawback is that the
precomputation phase must be re-executed each time the Sun zenith angle changes.
The Haber model also supports multiple scattering, with less approximations than
in the Nishita96 model. This increases its accuracy compared to Nishita96, at
the price of longer precomputations and less supported viewpoints. Its main
drawbacks are that the precomputation phase must be re-executed each time the
Sun zenith angle changes (as in Nishita96) and the approximation of the Rayleigh
phase function with the isotropic phase function, which limits its accuracy. The
Bruneton model does not use this approximation to compute multiple scattering,
and precomputes data for all Sun zenith angles. This increases its accuracy
compared to the Haber model, at the price of increased memory needs. The Elek
model uses more wavelengths, which increases its accuracy compared to the
Bruneton model (\cf the next section), at the cost of a longer precomputation
phase.

\newcommand{\rgbrenderingcell}[6]{
\begin{overpic}[width=0.198\columnwidth]{image_#3_#2_#1}
\ifthenelse{\equal{#1}{measurements}}{
}{
  \put(5,5){\footnotesize\color{white}%
    \input{program/output/figures/error_#4_#2_#1.txt}}
}
\ifthenelse{\equal{#5}{true}}{
  \put(82,85){\footnotesize\color{white}#6}
}{
}
\end{overpic}}

\newcommand{\rgbrenderingrow}[4]{
\rgbrenderingcell{nishita93}{#1}{#2}{#3}{#4}{\hspace{1.4mm}3} &
\rgbrenderingcell{nishita96}{#1}{#2}{#3}{#4}{\hspace{1.4mm}3} &
\rgbrenderingcell{preetham}{#1}{#2}{#3}{#4}{\hspace{1.4mm}3} &
\rgbrenderingcell{oneal}{#1}{#2}{#3}{#4}{\hspace{1.4mm}3} &
\rgbrenderingcell{haber}{#1}{#2}{#3}{#4}{\hspace{1.4mm}8} &
\rgbrenderingcell{bruneton}{#1}{#2}{#3}{#4}{\hspace{1.4mm}3} &
\rgbrenderingcell{elek}{#1}{#2}{#3}{#4}{15} &
\rgbrenderingcell{hosek}{#1}{#2}{#3}{#4}{11} &
\rgbrenderingcell{libradtran}{#1}{#2}{#3}{#4}{40} &
\rgbrenderingcell{measurements}{#1}{#2}{#3}{#4}{40}}

\newcommand{\rgbdiffcell}[2]{
\begin{overpic}[width=0.198\columnwidth]{image_approx_diff_#2_#1}
\put(5,5){\footnotesize\color{white}%
  \input{program/output/figures/image_approx_psnr_#2_#1.txt}dB}
\end{overpic}}

\newcommand{\rgbdiffrow}[1]{

\begin{overpic}[width=0.198\columnwidth]{image_approx_diff_#1_nishita93}
\put(5,5){\footnotesize\color{white}%
  \input{program/output/figures/image_approx_psnr_#1_nishita93.txt}dB}
\end{overpic} & 

\begin{overpic}[width=0.198\columnwidth]{image_approx_diff_#1_nishita96}
\put(5,5){\footnotesize\color{white}%
  \input{program/output/figures/image_approx_psnr_#1_nishita96.txt}dB}
\end{overpic} &

\begin{overpic}[width=0.198\columnwidth]{image_approx_diff_#1_preetham}
\put(5,5){\footnotesize\color{white}%
  \input{program/output/figures/image_approx_psnr_#1_preetham.txt}dB}
\end{overpic} &

\begin{overpic}[width=0.198\columnwidth]{image_approx_diff_#1_oneal}
\put(5,5){\footnotesize\color{white}%
  \input{program/output/figures/image_approx_psnr_#1_oneal.txt}dB}
\end{overpic} &

\begin{overpic}[width=0.198\columnwidth]{image_approx_diff_#1_haber}
\put(5,5){\footnotesize\color{white}%
  \input{program/output/figures/image_approx_psnr_#1_haber.txt}dB}
\end{overpic} &

\begin{overpic}[width=0.198\columnwidth]{image_approx_diff_#1_bruneton}
\put(5,5){\footnotesize\color{white}%
  \input{program/output/figures/image_approx_psnr_#1_bruneton.txt}dB}
\end{overpic} &

\begin{overpic}[width=0.198\columnwidth]{image_approx_diff_#1_elek}
\put(5,5){\footnotesize\color{white}%
  \input{program/output/figures/image_approx_psnr_#1_elek.txt}dB}
\end{overpic} &

\begin{overpic}[width=0.198\columnwidth]{image_approx_diff_#1_hosek}
\put(5,5){\footnotesize\color{white}%
  \input{program/output/figures/image_approx_psnr_#1_hosek.txt}dB}
\end{overpic} &

\begin{overpic}[width=0.198\columnwidth]{image_approx_diff_#1_libradtran}
\put(5,5){\footnotesize\color{white}%
  \input{program/output/figures/image_approx_psnr_#1_libradtran.txt}dB}
\end{overpic} &

\begin{overpic}[width=0.198\columnwidth]{image_approx_diff_#1_measurements}
\put(5,5){\footnotesize\color{white}%
  \input{program/output/figures/image_approx_psnr_#1_measurements.txt}dB}
\end{overpic}}

\begin{figure*}
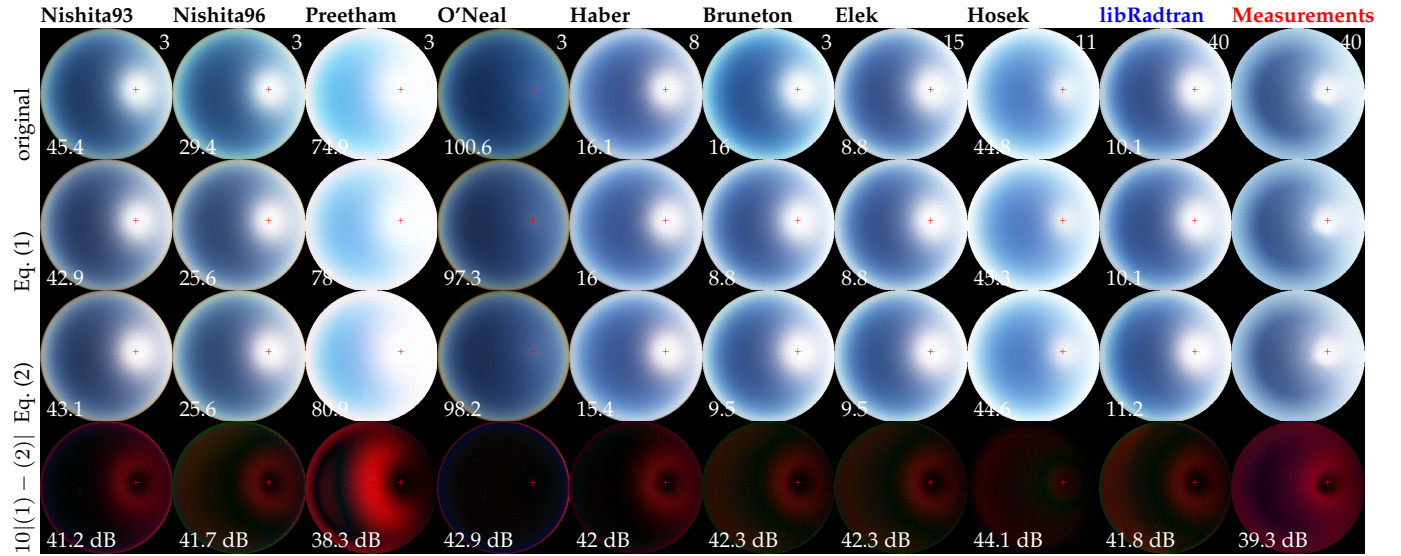

\begin{center}
\begin{mytabular}{p{0.04\columnwidth}*{10}{p{0.198\columnwidth}}}
&
\footnotesize\bfseries Nishita93 &
\footnotesize\bfseries Nishita96 &
\footnotesize\bfseries Preetham &
\footnotesize\bfseries O'Neal &
\footnotesize\bfseries Haber &
\footnotesize\bfseries Bruneton &
\footnotesize\bfseries Elek &
\footnotesize\bfseries Hosek &
\footnotesize\bfseries\leavevmode\color{blue} libRadtran &
\footnotesize\bfseries\leavevmode\color{red} Measurements\vspace{1mm}\\
\rotatebox{90}{\footnotesize original} & 
\rgbrenderingrow{10h15}{original}{original_rgb}{true}\\
\rotatebox{90}{\footnotesize \cref{eq:spectraltorgb}} & 
\rgbrenderingrow{10h15}{full_spectral}{rgb}{false}\\
\rotatebox{90}{\footnotesize \cref{eq:rgb}} & 
\rgbrenderingrow{10h15}{approx_spectral}{approximate_rgb}{false}\\
\rotatebox{90}{\footnotesize $10|\mathrm{(1)}-\mathrm{(2)}|$} &
\rgbdiffrow{10h15}
\end{mytabular}
\end{center}
\caption{{\bf Spectrum sampling}. Rendering of the skydome by using the number
of wavelengths $n_\lambda$ proposed by the authors of each model (1st row,
$n_\lambda$ in the top right), by using $n_\lambda=40$ and
\cref{eq:spectraltorgb} (2nd row), and by using $n_\lambda=3$ and \cref{eq:rgb}
(3rd row). The bottom left number in each cell is the RMSE, times 1000, compared
to the top right cell (summed over the 81 measurement samples, on the RGB colors
tone-mapped to the $[0,1]$ interval). The 4th row is the difference between the
2nd and 3rd ones, times 10, with the corresponding peak signal to noise ratio
(PSNR).}\label{fig:rgb_rendering}
\end{figure*}

Finally, an important conclusion from the above comparisons is that none of the
models is "perfectly" accurate, and that there is still room for improvement in
this area (although users do not necessarily prefer the most accurate models --
\cf \cref{sec_study}). Some ideas to increase accuracy include using more
wavelengths, taking polarization into account, or modeling aerosol optical
properties more precisely. In the next sections we show that only the latter
could potentially significantly increase accuracy.

\subsection{Spectrum sampling}\label{sec_rgborspectral}

In sections~\ref{sec_nishita93} to \ref{sec_hosek} we evaluated the models by
using 40 wavelengths between $360\,nm$ and $830\,nm$ as in\cite{Kider14},
although the models originally use between 3 and 15 wavelengths. Here we discuss
the impact of the number of wavelength samples used on the accuracy of the
results. We show that it is possible to use only 3 samples to reconstruct RGB
colors or full spectrums, with almost the same accuracy as with 40 samples.

\subsubsection*{RGB rendering}\label{sec_rgbrender}

In order to evaluate the impact of the number of wavelengths $n_\lambda$ used to
compute the final RGB color, we ran each model with the original number of
wavelengths proposed by their authors (or 3 if not specified). Also, for the
special case $n_\lambda=3$, we used the 3 spectral radiance samples directly as
RGB components\footnote{We also replaced the solar spectrum with a constant
function, i.e. we used $S(\lambda)=\mathrm{cste}$, as in the authors
implementation of the O'Neal and Bruneton models.}, instead of converting them
to RGB via the CIE color matching functions (which, as pointed out
in\cite{Elek10}, is incorrect -- but is done in the authors implementation of
the O'Neal and Bruneton models).

The results do not show significant differences between
$n_\lambda=40,\ 15,\ 11\ \mathrm{or}\ 8$ (see \cref{fig:rgb_rendering} and the
supplemental material). However, they clearly show that using a constant solar
spectrum and 3 spectral radiance samples directly as RGB components
significantly decreases the accuracy\footnote{Strangely, for the morning scene
in our perceptual study, participants find this less accurate method
significantly more realistic (\cf the results for the Elek and Bruneton models
in \cref{tab:prefmatrix}). We think this is due to the constant solar spectrum
approximation, which gives an (unintended) "white balance" effect. Indeed, if we
add a "white balance" effect to the Elek result (by dividing each pixel by the
color corresponding to the extraterrestrial solar spectrum), participants find
this new image significantly more realistic than the original one (and as
realistic as the Bruneton image). This suggests that, besides physical accuracy,
it is also important to correctly simulate perceptual effects (such as tone
mapping and white balance) to improve realism.}. In the following we propose a
sound and more accurate method to convert these 3 samples to an RGB color.

With the reference spectral method the linear sRGB components are given by
\begin{equation}
[c_r,c_g,c_b]=\int_{\lambda_\mathrm{min}}^{\lambda_\mathrm{max}}
[\tilde{r}(\lambda),\tilde{g}(\lambda),\tilde{b}(\lambda)]
L(\lambda)\mathrm{d}\lambda\label{eq:spectraltorgb}
\end{equation}
where $[\tilde{r},\tilde{g},\tilde{b}]^\top=M[\bar{x},\bar{y},\bar{z}]^\top$ are
color matching functions computed from the XYZ color matching functions
$\bar{x},\bar{y},\bar{z}$ and from the XYZ to linear sRGB conversion matrix $M$.
Using our knowledge that the sky spectral radiance $L(\lambda)$ is proportional
to the extraterrestrial solar spectrum $S(\lambda)$ and to $\lambda^{-4}$
without aerosols, we can approximate $L(\lambda)$ near some $\lambda_0$ with
$L(\lambda_0)S(\lambda)\lambda^{-\alpha}/S(\lambda_0){\lambda_0}^{-\alpha}$,
where $\alpha\le 4$ due to aerosols. Substituting this in
\cref{eq:spectraltorgb} for 3 wavelengths $\lambda_r, \lambda_g, \lambda_b$ we
get
\begin{equation}
[c_r,c_g,c_b]\approx
[k_r L(\lambda_r), k_g L(\lambda_g), k_b L(\lambda_b)]\label{eq:rgb}
\end{equation}
where $k_r=\int_{\lambda_\mathrm{min}}^{\lambda_\mathrm{max}} \tilde{r}(\lambda)
\left(S(\lambda)\lambda^{-\alpha}/S(\lambda_r){\lambda_r}^{-\alpha}\right)
\mathrm{d}\lambda$, and similarly for $k_g$ and $k_b$ (in our results we used
$\alpha=3$ and $\lambda_r,\lambda_g,\lambda_b=680,550,440\,nm$). In other words,
we can compute an RGB color by multiplying 3 spectrum samples
$L(\lambda_r),L(\lambda_g),L(\lambda_b)$ by 3 constants. And, as shown in
\cref{fig:rgb_rendering}, this approximation is almost as accurate as with 40
wavelengths (the PSNR is about $40\,\mathrm{dB}$, for all models).

\subsubsection*{Spectral rendering}

For applications that require the sky spectral radiance $L(\lambda)$ at
$n_\lambda$ wavelengths (\eg energy studies), we can either compute $L$ for each
wavelength, or we can compute $L$ for a smaller number of wavelengths, and
extrapolate its value for the other wavelengths, to save computations. We show
here that we can reconstruct a full spectrum from only 3 samples with a very
small approximation error.

A simple method to reconstruct $L(\lambda)$ from 3 samples
$L(\lambda_r),L(\lambda_g),L(\lambda_b)$ is to express it as a linear
combination of the 3 CIE D65 standard illuminant components \cite{CIE04}, as in
\cite{Preetham99}. This approximation is quite accurate, but we found that the
following method, slightly less simple, is much more accurate:
\begin{itemize}
\item convert $L(\lambda_r),L(\lambda_g),L(\lambda_b)$ to sRGB using
\cref{eq:rgb},
\item convert from sRGB to XYZ and then to xyY,
\item convert the xyY color to a full spectrum as in \cite{Preetham99}.
\end{itemize}
Our results, in \cref{fig:approx_spectral}, show that this approximation
increases the RMSE by less than $1\,mW/(m^2.sr.nm)$, compared to a full
computation with 40 wavelengths. 

\begin{table}
\renewcommand{\arraystretch}{1.3}
\caption{{\bf Spectrum sampling}. The RMSE in $mW/(m^2.sr.nm)$, summed over the
17 daytimes, 81 directions, and $\lambda$ from $360$ to $720\,nm$. {\em Left}:
spectral radiance computed separately for each wavelength. {\em Right}:
approximate spectrum reconstructed from 3 spectral radiance computations.}
\label{fig:approx_spectral}
\centering
\begin{tabular}{|r||c|c|c|c|c|c|c|c|}
\hline
\bfseries & \bfseries Full spectrum & \bfseries Reconstruction \\
\bfseries Model & \bfseries computation & \bfseries from 3 samples \\
\hline
Nishita93 & 26.6
 &
26.4
 \\
\hline
Nishita96 & 18.3
 &
19.2
 \\
\hline
Preetham & 88.1
 &
88.9
 \\
\hline
O'Neal & 49.5
 &
49.5
 \\
\hline
Haber & 14.7
 &
14.4
 \\
\hline
Bruneton & 11.3
 &
11.4
 \\
\hline
Elek & 11.3
 &
11.4
 \\
\hline
Hosek & 41.5
 &
39
 \\
\hline
\end{tabular}
\end{table}

\subsection{Polarization}

Another idea to increase accuracy is to take polarization into account.
Although the Sun light is initially not polarized, it becomes (partially)
linearly polarized after the first scattering event in the atmosphere. Because
of this, after two scattering events at $90^\circ$, the scattered intensity can
be null, a phenomenon that can't be simulated if polarization is ignored. Can we
then significantly increase the accuracy of the Computer Graphics clear sky
models by taking polarization into account?

\begin{figure}
\begin{center}
\begin{overpic}[width=0.25\columnwidth]{%
  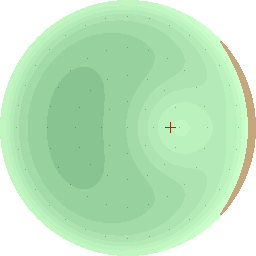}
\end{overpic}%
\begin{overpic}[width=0.25\columnwidth]{%
  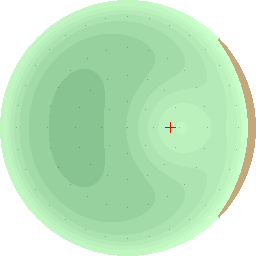}
\end{overpic}%
\begin{overpic}[width=0.25\columnwidth]{%
  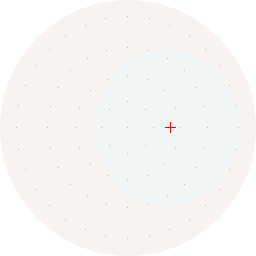}
\put(5,5){RMSE=\footnotesize%
  \input{program/output/figures/error_sza30_polradtran_scalar.txt}}
\end{overpic}%
\end{center}
\caption{{\bf Effect of polarization}. Absolute luminance with (left) or without
(middle) polarization, and the relative error between the two
(right).}\label{fig:polarization}
\end{figure}

To answer this we used the polradtran solver in libRadtran, which can take
polarization into account or not, and we compared the results obtained with or
without polarization, for the same atmosphere model (due to some polradtran
restrictions we couldn't use the atmosphere properties from
\cref{sec_model_parameters}. Instead, we used default atmosphere and aerosols
properties provided with libRadtran). Our results, in \cref{fig:polarization},
show that the difference between the two is very small, much smaller than the
difference between the CG models and the measurements. In other words, taking
polarization into account would not significantly increase their accuracy.

\subsection{Aerosol properties}

The Computer Graphics clear sky models use very few parameters for the aerosols.
The analytical models use only one parameter, the turbidity, while the others
generally use the aerosols scattering and absorption coefficients, their scale
height, and the asymmetry factor $g$ of their phase function. In contrast,
libRadtran can use any number of layers, where each layer can specify
wavelength-dependent parameters and arbitrary phase functions\footnote{the Haber
model also uses layers, but is still restricted to phase functions depending on
a single parameter $g$.}. We think adding more aerosol parameters in the CG
models is the best way to significantly increase their accuracy. Also, unlike
adding more wavelengths or supporting polarization, this would not significantly
decrease their performance.

\begin{figure}
\begin{center}
\includegraphics[width=0.9\columnwidth]{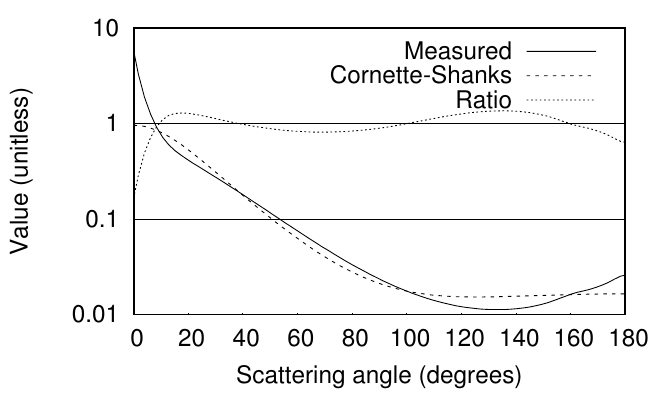}
\end{center}
\caption{{\bf Aerosols phase function}. Comparison between a typical measured
phase function for aerosols, and the Cornette-Shanks analytical model for the
corresponding asymmetry factor.}\label{fig:phasefunction}
\end{figure}

In particular, we think using more realistic phase functions for aerosols is the
easiest way to increase accuracy in the solar aureole region. Indeed, all the CG
models are quite inaccurate in this region, where single scattering due to
aerosols is the most significant term. And, comparing the
Cornette-Shanks\cite{TS99} phase function with a typical aerosol phase function
clearly shows that the analytical model is not really accurate for scattering
angles smaller than 20 degrees (see \cref{fig:phasefunction}).

\section{Conclusion}\label{sec_conclusion}

We have presented a qualitative and quantitative evaluation of 8 clear sky
models used in Computer Graphics. We compared the models with each other and
with a reference model from the physics community, as well as with ground truth
measurements, and via a perceptual study. All the models are based on the same
underlying physical equations, but make different hypothesis and approximations
to achieve different goals (\eg speed \vs accuracy). Our results show, as could
be expected, that the less simplifications and approximations are used to solve
the physical equations, the more physically accurate are the results. They also
show that accuracy can still be improved, and that the most promising way to
achieve this is to model aerosols more precisely (as opposed to, for instance,
simulating more wavelengths, or taking polarization into account).


%

\appendices


\ifCLASSOPTIONcompsoc
  \section*{Acknowledgments}
\else
  \section*{Acknowledgment}
\fi

We would like to thank Janne Kontkanen and Evan Parker for proofreading this
paper.

\ifCLASSOPTIONcaptionsoff
  \newpage
\fi



\bibliographystyle{IEEEtran}
\bibliography{IEEEabrv,paper}
\end{document}